\documentclass[acmsmall]{acmart}

\usepackage{caption}
\usepackage{subcaption}
\usepackage{array}
\usepackage{multirow}
\usepackage{enumitem}
\usepackage{booktabs}
\usepackage{soul}
\usepackage{balance}

\AtBeginDocument{%
  \providecommand\BibTeX{{%
    \normalfont B\kern-0.5em{\scshape i\kern-0.25em b}\kern-0.8em\TeX}}}

\copyrightyear{2023}
\acmYear{2023}
\setcopyright{acmlicensed}

\acmConference[CHI '23]{Proceedings of the 2023 CHI Conference on Human Factors in Computing Systems}{April 23--28, 2023}{Hamburg, Germany}
\acmBooktitle{Proceedings of the 2023 CHI Conference on Human Factors in Computing Systems (CHI '23), April 23--28, 2023, Hamburg, Germany}
\acmPrice{15.00}
\acmDOI{10.1145/3544548.3581570}
\acmISBN{978-1-4503-9421-5/23/04}

\begin{document}

\title[Thought Bubbles: A Proxy into Players' Mental Model Development]{Thought Bubbles: A Proxy into Players' Mental Model Development}

\author{Omid Mohaddesi}
\orcid{0000-0002-6245-4003}
\affiliation{%
  \institution{Northeastern University}
  \streetaddress{360 Huntington Ave.}
  \city{Boston}
  \state{Massachusetts}
  \country{USA}
  \postcode{02115}
}
\email{mohaddesi.s@northeastern.edu}

\author{Noah Chicoine}
\orcid{0000-0002-0602-1086}
\affiliation{%
  \institution{Northeastern University}
  \streetaddress{360 Huntington Ave.}
  \city{Boston}
  \state{Massachusetts}
  \country{USA}
  \postcode{02115}
}
\email{chicoine.n@northeastern.edu}

\author{Min Gong}
\orcid{0000-0002-8686-3442}
\affiliation{%
  \institution{Northeastern University}
  \streetaddress{360 Huntington Ave.}
  \city{Boston}
  \state{Massachusetts}
  \country{USA}
  \postcode{02115}
}
\email{gong.mi@northeastern.edu}

\author{Ozlem Ergun}
\affiliation{%
  \institution{Northeastern University}
  \streetaddress{360 Huntington Ave.}
  \city{Boston}
  \state{Massachusetts}
  \country{USA}
  \postcode{02115}
}
\email{o.ergun@northeastern.edu}

\author{Jacqueline Griffin}
\affiliation{%
  \institution{Northeastern University}
  \streetaddress{360 Huntington Ave.}
  \city{Boston}
  \state{Massachusetts}
  \country{USA}
  \postcode{02115}
}
\email{ja.griffin@northeastern.edu}

\author{David Kaeli}
\affiliation{%
  \institution{Northeastern University}
  \streetaddress{360 Huntington Ave.}
  \city{Boston}
  \state{Massachusetts}
  \country{USA}
  \postcode{02115}
}
\email{d.kaeli@northeastern.edu}

\author{Stacy Marsella}
\affiliation{%
  \institution{Northeastern University}
  \streetaddress{360 Huntington Ave.}
  \city{Boston}
  \state{Massachusetts}
  \country{USA}
  \postcode{02115}
}
\email{s.marsella@northeastern.edu}

\author{Casper Harteveld}
\affiliation{%
  \institution{Northeastern University}
  \streetaddress{360 Huntington Ave.}
  \city{Boston}
  \state{Massachusetts}
  \country{USA}
  \postcode{02115}
}
\email{c.harteveld@northeastern.edu}

\renewcommand{\shortauthors}{Mohaddesi et al.}

\begin{abstract}
Studying mental models has recently received more attention, aiming to understand the cognitive aspects of human-computer interaction. However, there is not enough research on the elicitation of mental models in complex dynamic systems. We present \emph{Thought Bubbles} as an approach for eliciting mental models and an avenue for understanding players' mental model development in interactive virtual environments. We demonstrate the use of Thought Bubbles in two experimental studies involving 250 participants playing a supply chain game. In our analyses, we rely on Situation Awareness (SA) levels, including perception, comprehension, and projection, and show how experimental manipulations such as disruptions and information sharing shape players' mental models and drive their decisions depending on their behavioral profile. Our results provide evidence for the use of thought bubbles in uncovering cognitive aspects of behavior by indicating how disruption location and availability of information affect people's mental model development and influence their decisions.
\end{abstract}



\maketitle

\newcommand{\newtext}[1]{#1}

\newcommand{\oldtext}[1]{}

\newcommand{\newtextfinal}[1]{#1}

\newcommand{\oldtextfinal}[1]{}

\section{Introduction}

\oldtext{Many complex dynamic systems, such as supply chains, involve human decision-makers at their focal point.}\newtext{Complex dynamic systems, such as supply chains, comprise a network of interdependent actors that constantly interact with each other~\cite{ridolfi2012complex}. The behavior of these actors results in emergent dynamics and influences how they interact with the system over time~\cite{bekebrede2015understanding, lukosch2018scientific}. Some actors show distinct}\oldtext{Human behavior is highly influenced by how humans interact with the system, resulting in varying} decision patterns~\cite{mohaddesi_trust_2022}\newtext{, and} some\oldtext{decision-makers may} even exhibit irrational behavior, such as panic-buying and hoarding, in the context of supply chains~\cite{sterman2015m}. \newtext{Such behaviors and varying decisions have been shown to be}\oldtext{Prior work provides evidence that distinct decision patterns are} the outcome of different mental models about the underlying system~\cite{gary2011mental, brown2009role}. 
Therefore, it is vital to investigate how decision-makers form mental models of such complex dynamic systems \newtext{over time and undisturbed in the decision context. 
}%
\oldtext{to augment decision-making.} Such investigation, however, calls for methods for eliciting mental models and evaluating mental model \oldtext{formation}\newtext{development over time}~\cite{van2014system}.

\oldtext{Studying mental models is of particular interest to the HCI field. However, most research on mental models in HCI is focused on users' understanding of some artifacts such as game aspects
, AI agents
, game avatars
, or interaction with artifacts through gestures
. These efforts have primarily focused on improving the design by understanding mental models
, or improving users' understanding of how the artifact works to improve performance
. In addition, prior studies mostly rely on a non-diegetic elicitation of mental models, meaning that they query users out of context (e.g., before or after their interaction). Little research exists on mental model formation during the interaction with dynamic virtual environments that can elicit mental models in a diegetic manner.}


\newtext{In this paper, we present such an elicitation method that we refer to as \textit{thought bubbles}. This method evokes thought processes in situ and diegetically through multiple open-ended prompts over time as part of an interactive virtual environment, and thus seeks to elicit \textit{mental model development} in the context of complex dynamic systems. We use the term mental model development to refer to a temporal process that happens during the interaction with complex dynamic systems and over time.} The need for \newtext{a process view}\oldtext{diegetic elicitation} \newtextfinal{stems}\oldtextfinal{arises} from\oldtextfinal{the fact that} mental models \oldtextfinal{are}not \newtextfinal{being} static constructs~\cite{landriscina2013simulation, crandall2006working}. They dynamically change depending on new information or perceptions of the system. Previous studies provide evidence that the interaction forms behavior~\cite{mohaddesi_trust_2022}, and that behavior is the outcome of cognitive constructs~\cite{klein1993decision, endsley2015situation}. Hence, it is not unreasonable to think that the interaction affects mental model development. \newtext{In addition, we leverage interactive virtual environments because they have been demonstrated to be useful for studying and learning about complex dynamic systems~\cite{lukosch2018scientific, mohaddesi_introducing_2020,bekebrede2006build} and allow for eliciting mental models diegetically, which mitigates the effect of any out-of-context effects such as pre/post-interaction information or experimenter bias~\cite{doyle1998mental}.} 


\oldtext{We use a gamette environment as described by Mohaddesi et al.
and conduct two experimental studies (Study 1: $n=115$ and Study 2: $n=135$) in a supply chain context to test thought bubbles for eliciting players' mental model development.}Eliciting mental models is only the first step in gaining insight into the cognitive aspect of human actions~\cite{grenier2015conceptual}. We also need reliable methods for analyzing the elicited mental model concepts. As our elicited mental models result from textual verbalization of thought processes, we can leverage qualitative analysis to make sense of our elicitation~\cite{gero2020mental, jones2011mental, grenier2015conceptual}. In grounding our qualitative analysis, we must align our outcomes with existing theories to gain a theoretical understanding of mental model development. We \newtext{rely on}~\oldtext{found} the Situation Awareness (SA) model, introduced by Endsley~\cite{endsley1995toward}, \newtext{as} a~\oldtext{useful} guiding framework for our qualitative analysis. SA framework is a well-established model that describes people's \textit{perception} of the elements of the environment, the \textit{comprehension} of their meaning, and the \textit{projection} of future states~\cite{endsley1988situation}; hence providing a window into their mental model \newtext{development}\oldtext{formation}~\cite{endsley2000situation}. \oldtext{We test the reliability of this approach by using the results of our qualitative coding in Study 1 for the qualitative coding of Study 2. Finally, we complement our qualitative results with quantitative analysis and show how players form mental models of a disrupted supply chain depending on disruption location, information sharing, and their observed behavioral profile.}

\newtext{We test thought bubbles and our qualitative analysis framework using an interactive virtual environment called \textit{gamettes}~\cite{mohaddesi_introducing_2020} and conduct two experimental studies (Study 1: \textit{n}=$115$ and Study 2: \textit{n}=$135$) in a supply chain context. These studies are exploratory in nature and serve two purposes. First, we test whether elicitation via thought bubbles can offer us a meaningful outlook on the cognitive aspects of decision-making. In both studies, we examine how experimental manipulations such as disruption location in the supply chain or the level of information sharing affect mental model development. Study 2 also investigates mental model development to help explain cognitive aspects of decisions made by players with different behavioral profiles. Second, we use the results of our qualitative coding in Study 1 as a codebook for the qualitative analysis of Study 2, aiming to test the reliability of SA for analyzing elicited mental model concepts over time.} The following are the contributions of this work:

\begin{itemize}
    \item We present \textit{thought bubbles} as a method for collecting qualitative data on human thought processes diegetically and temporally to elicit mental model development;
    \item We show a mixed-method approach for analyzing elicited mental models to gain a theoretical understanding of cognitive aspects of human decisions and their mental model formation; and
    \item We provide evidence on the effect of disruption location and information sharing on mental model development (Study 1), as well as the effect of information sharing on the mental model \oldtext{formation}~\newtext{development} of players with distinct behavioral profiles (Study 2).
\end{itemize}

\section{Related Work}
In our review of the related literature, we first study the notion of mental models as a cognitive construct and methods for eliciting mental models. We then look into the study of mental models in the context of dynamic decision-making.

\subsection{Mental Models}\label{RW:Mental_Models}
Mental models are internal and small-scale representations of the real world that people rely on in their interaction with external reality~\cite{jones2011mental, gentner2014mental}. The notion of mental models, which was first put forth in 1943 by Craik~\cite{craik1943nature}, has been extensively researched by cognitive psychologists and tested in various contexts. As a cognitive construct, mental models are compared with the notion of ``schemata'', which are sets of expectations or long-term knowledge structures based on prior experiences~\cite{brewer1981role, wilson1989mental,johnson1983mental,rumelhart1984schemata}. Psychologists typically refer to both concepts to better contrast their definitions~\cite{holland1989induction, brewer1987schemas, rutherford1991models}. The difference is that schemata are inflexible and generic knowledge structures that provide predictive knowledge for routine situations~\cite{holland1989induction, rutherford1991models}. Mental models, on the other hand, are dynamically formed each time a new data or situation is perceived, and people form richer and more consistent mental models as they gain more experience~\cite{crandall2006working}. When faced with unfamiliar domains, people may rely on analogies to extend knowledge from prior experiences that are well-understood and perceived as similar~\cite{gentner2014mental,collins1987people}. 

Prior research also describes how mental models can be used to run mental simulations (i.e., mentally project into the future) to envision and explore alternative future states before acting~\cite{crandall2006working,landriscina2013simulation}. This is where cognition meets behavior; the mental model construct directs actions when people rely on mental simulations in interacting with the environment~\cite{klein1993decision}. From this perspective, understanding behavior requires gaining insight into mental models. However, we cannot directly observe cognitive constructs such as mental models. We need instruments and methodologies to allow people to externalize their thought process~\cite{ifenthaler2010relational}. With thought bubbles, we aim to gain insight into the cognitive aspect of human decision-making by eliciting mental models. 

Prior research points to the lack of consistent methodologies for mental model elicitation~\cite{grenier2015conceptual}. The literature refers to elicitation approaches in two dimensions: (1) techniques used for extracting thought processes (i.e., verbal, visual, or hybrid); and (2) elicitation context (i.e., situated vs. non-situated)~\cite{jones2014eliciting}. Verbal elicitation refers to using dialogue or discussion in an interview procedure. In contrast, visual elicitation is related to using graphics to represent an individual's mental model. The hybrid approach concerns a combination of verbal and visual elicitation~\cite{grenier2015conceptual, cooke1994varieties, lamere2020making}. Situated practices involve eliciting mental models in the context (i.e., diegetic). The non-situated approach, on the other hand, involves elicitation in a location removed from the topic of study~\cite{jones2014eliciting}. Thought bubbles are a form of verbal elicitation that we utilize in situ. By allowing players to provide open-response comments, we leverage the verbalization of players' thought processes to investigate their mental model development in the context of dynamic decision-making and without introducing framing bias~\cite{tversky1985framing,memon2013enhanced}. 





\subsection{Dynamic Decision Making in Interactive Systems}\label{RW:Dynamic_Decision_Making}
\newtext{Dynamic decision-making tasks involve taking a series of actions in a dynamic system over time to achieve some goal (e.g., maximizing total profit)~\cite{edwards1962dynamic}. Actions are interdependent, which means later actions depend on the earlier ones, and the environment will change spontaneously or as the result of earlier actions~\cite{hotaling2015dynamic, klein1993decision}.} The study of mental models is fundamental in dynamic decision-making environments\oldtext{ where the situation changes from time to time or in response to human actions
}. Most studies in this area are motivated by trying to develop a better understanding of mental models to improve decision-making~\cite{doyle1998mental}. For example, researchers have studied mental model elicitation from experts to improve system dynamics models for mental model improvement~\cite{ford1998expert}. Others have studied individuals' understanding of the decision context and showed how mental models affect human rationality in dynamic decisions~\cite{rouwette2004exploring}, how different mental models lead to different decisions and performances~\cite{gary2011mental}, and 
how higher similarity between mental models and the decision context results in higher performance~\cite{ritchie2001informing}. Most of these studies rely on a simulation for the decision context and a form of elicitation technique, as described in the previous section, to capture mental models. 

~\oldtext{Researchers have also proposed theories for analyzing elicitation results in dynamic decision-making. For example, Andersen and Rohrbaugh
examined mental models by linking system dynamics with models of judgment by studying three main sub-models: ends model, means model, and ends-means model. Richardson et al.
pointed to this framework's shortcoming and argued that mental models cannot be elicited without distortion. Perhaps people might not be able to articulate all aspects of their knowledge or fail to verbalize what seems obvious
. While we agree that distortion is inevitable, we show the value of thought bubbles in approximating players' thought processes by studying mental model development over time. We found the Situation Awareness (SA) model introduced by Endsley
suitable for approximating mental model development.}
~\newtext{Researchers have proposed various approaches for analyzing elicitation results and studying mental models. From our research perspective, these approaches fall into one of three categories: (1) not applicable to dynamic decision-making, (2) applicable but not practical (e.g., demonstrated by lack of empirical use), and (3) applicable and practical. As an example of the first, Johnson-Laird~\cite{johnson1980mental} studied mental models in connection with human reasoning though propositional reasoning, deduction, and syllogistic inference~\cite{johnson1984syllogistic,johnson1992propositional,johnson1983mental}. Despite its prominence, Johnson-Laird's view is not immediately applicable to dynamic decision-making environments because of its lack of support for studying mental models formed through interaction with the task or the environment~\cite{staggers1993mental, brown2009role}. 
As for the second, Richardson et al.~\cite{richardson1994foundations} developed a theory of perception, planning, action, and learning specifically for studying mental models in dynamic decision-making. They followed the idea that mental models are multifaceted (i.e., comprising three main sub-models: ends model, means model, and ends-means model~\cite{andersen1992some}) and linked system dynamics feedback theory with models of judgment. Unfortunately, there is not enough evidence in the literature on the empirical use of the theory proposed by Richardson et al.~\cite{richardson1994foundations}. 

Finally, as for the third, Endsley~\cite{endsley1995toward} introduced the theory of Situation Awareness (SA) in dynamic decision-making and proposed its use for studying mental models~\cite{endsley2000situation}. SA model has been extensively and empirically researched through numerous studies by Endsley~\cite{endsley2015situation, endsley2018expertise, endsley2017autonomous, endsley2021systematic} and others~\cite{chen2021developing, teichmann2021situation, andrews2022role}. Therefore, we rely on the SA model for approximating players' mental model development over time. This idea of leveraging the use of SA to characterize mental models has also been examined by other researchers in inductive reasoning tasks~\cite{zhang2022using}.} Endsley describes situation awareness in dynamic decision-making as: 

\begin{quote}
``[...] the perception of the elements in the environment within a volume of time and space, the comprehension of their meaning, and the projection of their status in the near future.''~\cite{endsley1988situation}
\end{quote}

According to this model, situation awareness is achieved through: (1) perceiving the status, attributes, and dynamics of elements of the environment (perception); (2) understanding the situation based on a synthesis of perceived elements and determining their relevance to decision goals (comprehension); and (3) the ability to project the future events and dynamics of the elements of the system (projection). According to Endsley~\cite{endsley2000situation}, a situation model~\footnote{Here we use Situation Awareness and Situation Model interchangeably.} provides a useful window into people's mental models. While people may have a mental model of how a dynamic system (here a supply chain) works, their interaction with the environment directs how they update their situation model, which eventually directs the selection and revision of their mental model~\cite{endsley2015situation}. More specifically, mental models direct attention to key features of the environment and the comprehension, projections, and expectations. Therefore, a situation model is a key to understanding the mental model~\cite{endsley2000situation}.

Endsley~\cite{endsley1988situation,endsley1995toward} also introduced the Situation Awareness Global Assessment Technique (SAGAT) to measure individuals' situation awareness. SAGAT involves stopping the simulation in random intervals and querying the subjects with specific questions to determine their SA at that particular time~\cite{endsley1988situation}. \newtext{While we do not use SAGAT, with thought bubbles, we}~\oldtext{Thought bubbles} follow the same idea by prompting players in intervals. Our goal~\oldtext{, however,} is not to measure SA using thought bubbles. Instead, we use the SA framework for making sense of players' comments during our qualitative coding process \newtext{resulting from responding to the thought bubbles deployed in gamettes}. \oldtext{Therefore, we keep our prompt open-ended to evoke players thinking and allow them to verbalize their thought processes in an open-response format. This idea of leveraging the use of SA to characterize mental models has also been examined by other researchers in inductive reasoning tasks
. However, little is known about utilizing such a framework for making sense of cognitive aspects of human decision-making in dynamic systems. Here, we use a gamette
, a virtual interactive environment, to simulate dynamic decision-making in the supply chain context (Section
). We design thought bubbles as part of this environment and attempt to collect qualitative data on players' thought processes (Section
). Finally, we showcase how to leverage mixed-methods research to provide insight into mental model development (Section
and Section
).}

\section{Gamettes}\label{gamette}
\begin{figure}[ht]
    \centering
    \includegraphics[width=.5\columnwidth]{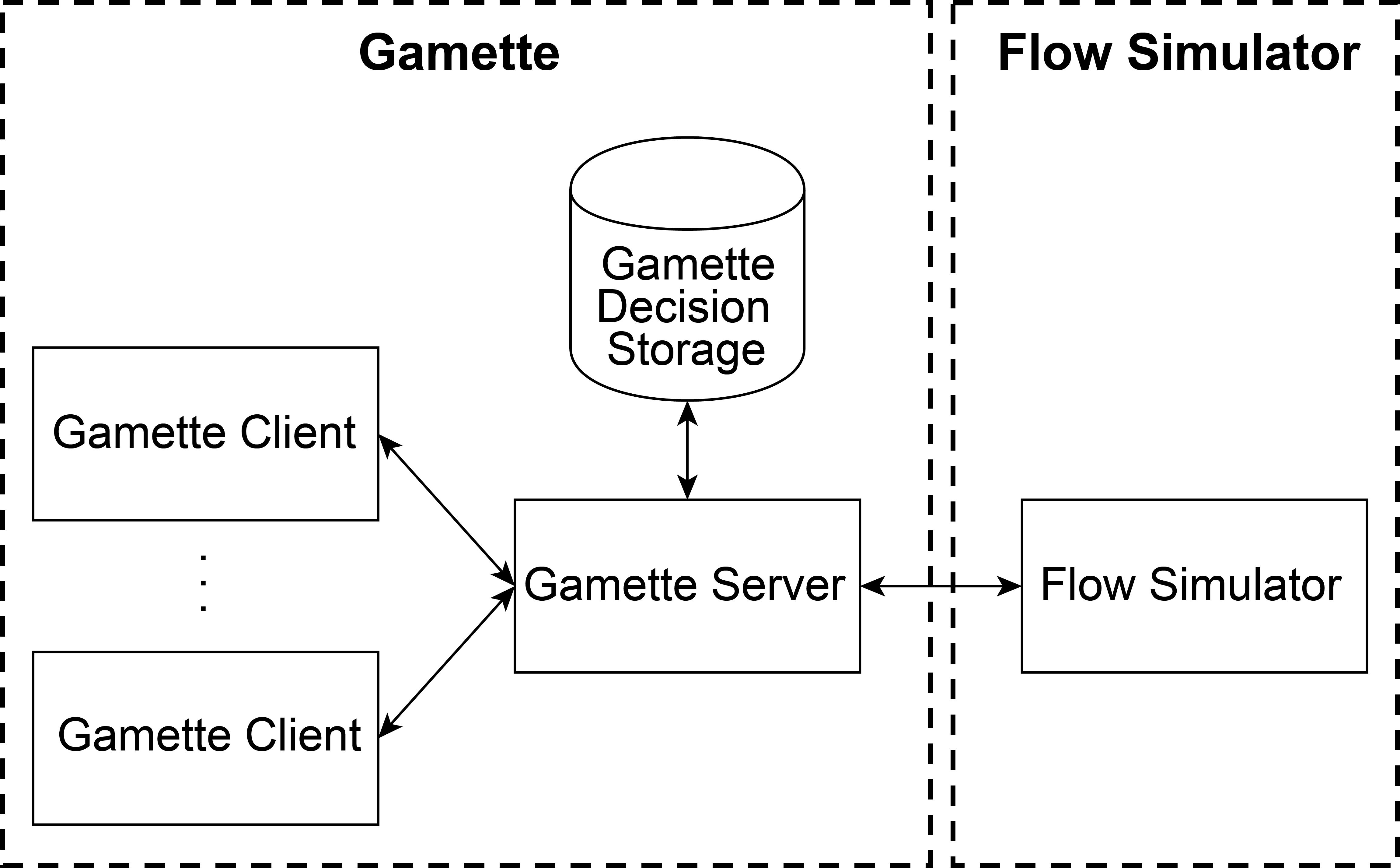}
    \caption{The integrated simulation framework
    .}\label{fig:simulationframework}
    \Description[Simulation architecture, including gamette and Flow Simulator]{The architecture of the integrated simulation framework shows how the gamette environment is connected to the Flow simulator. Each gamette environment comprises a gamette server that communicates with the Flow Simulator. Multiple gamette clients can communicate with the gamette server. Gamette server stores all game data in gamette decision storage.}
\end{figure}

\newtextfinal{We }\oldtextfinal{Mohaddesi et al.}
introduced gamettes \newtextfinal{in our previous study~\cite{mohaddesi_introducing_2020},} as a serious game approach for collecting data on behavioral aspects of human decision-making in supply chain experiments. A gamette is a short game-based scenario that immerses human decision-makers into a specific situation, requiring them to make decisions by responding to a dialog or taking actions. The term gamette is a contraction of ``game'' with ``vignette'', and similar to a vignette, a gamette aims to provide a \textit{brief description} of a situation, as well as to \textit{portray} someone. Here, we utilize gamettes to understand an individual's mental model supporting their decision-making in a drug delivery supply chain. For this, we use the integrated simulation framework proposed by Doroudi et al.~\cite{doroudi2018integrated}. This framework comprises a \textit{Flow Simulator} for simulating the supply chain dynamics and a \textit{gamette} environment for engaging human decision-makers with the simulation by immersing them in a specific role and particular state of the supply chain (see Figure~\ref{fig:simulationframework}).

Within this framework, the Flow Simulator is a multi-agent simulation and the central hub that controls the dynamics of a drug delivery supply chain, including the flow of information and physical products over time. These information and physical flows are driven by the decisions and actions taken by the agents of the system (i.e., manufacturers, wholesalers, and health centers). The Flow Simulator can run in a \textit{standalone mode} and without any human agents, which means that the Flow Simulator simulates the evolution of the supply chain system and also controls the decisions and actions of the agents through predefined policies. Alternatively, the Flow Simulator can simulate the evolution of the supply chain by fetching information from a gamette client that captures the decisions of human players. 

Following the approach \newtextfinal{in our prior work}\oldtextfinal{by Mohaddesi et al.}~\cite{mohaddesi_introducing_2020}, we created a gamette with StudyCrafter\footnote{StudyCrafter is a platform where users can easily create, play, and share gamified projects. It can be accessed at \url{https://studycrafter.com}}, where players take the role of a wholesaler in a drug delivery supply chain. Details of the gamette design are described in Section~\ref{study1}. The same gamette is used for Study 1 and Study 2; the difference is in their experimental design (i.e., Study 1 considers the disruption location and various forms of information sharing; Study 2 considers different behavioral profiles for decision-makers and how they respond to information sharing). 

\begin{figure*}[ht]
    \centering
    \includegraphics[width=\textwidth]{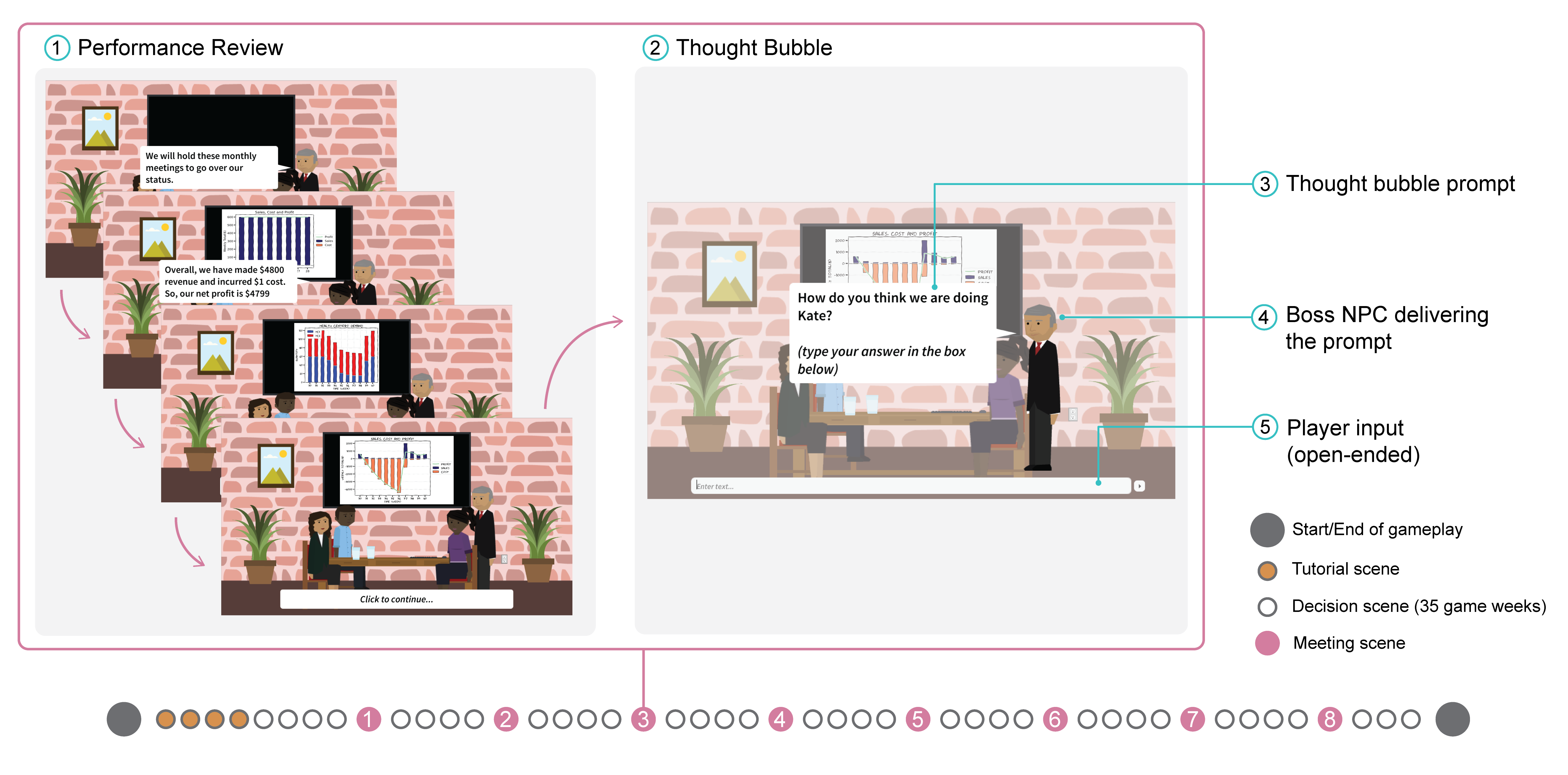}
    \caption{\newtext{Thought bubble design in the gamette as part of the meeting scene~\includegraphics[scale=0.75, trim=0 0.08cm 0 0]{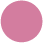}. Players experience the meeting scene every four weeks during the gameplay (8 times in total), starting on Week 24. Each time, players are first provided with a performance review~\includegraphics[scale=0.75, trim=0 0.08cm 0 0]{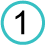} and then face the thought bubbles~\includegraphics[scale=0.75, trim=0 0.08cm 0 0]{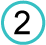}. An NPC delivers the thought bubble prompt~\includegraphics[scale=0.75, trim=0 0.08cm 0 0]{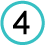} in an open-ended format to evoke players to reflect on their performance by asking ``How do you think we are doing Kate?''~\includegraphics[scale=0.75, trim=0 0.08cm 0 0]{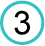}. Players provide their open-ended responses in the input area~\includegraphics[scale=0.75, trim=0 0.08cm 0 0]{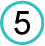}. Note that the gameplay phase starts on Week 21 after the tutorial ends.}}
    \label{fig:thoughtbubbles}
    \Description[Thought Bubble design]{Design of thought bubbles and its characteristics where players review performance and face the thought bubbles prompt. The prompt is delivered by NPC that asks, "How do you think we are doing Kate?" to evoke players to reflect on their performance. Players can provide their response in an open-ended format in an input area.}
\end{figure*}

\section{Thought Bubbles}\label{thought_bubbles}
\newtext{We designed thought bubbles with three main characteristics: (1) diegetic, (2) verbal and open-ended input, and (3) over time. Figure~\ref{fig:thoughtbubbles} demonstrates the design of thought bubbles and its characteristics. To ensure diegetic data collection, we included}\oldtext{We designed} thought bubbles \newtext{(see~\includegraphics[scale=0.75, trim=0 0.08cm 0 0]{figures/circle2.png} in Figure~\ref{fig:thoughtbubbles})} as part of a recurrent meeting scene in the gamette \oldtext{to collect data on players' mental model formation by prompting them to reflect on their task.}\newtext{where players would first review their performance 
(see~\includegraphics[scale=0.75, trim=0 0.08cm 0 0]{figures/circle1.png} in Figure~\ref{fig:thoughtbubbles}).} \oldtext{Figure
demonstrates a snapshot of this meeting scene in the gamette where the thought bubble is presented to players.}\oldtext{We designed this meeting scene to provide players with feedback on their performance during the \textit{gameplay} phase.}This feedback is provided to players through interaction with other \newtext{Non-Player Characters (}NPCs\newtext{)} \newtext{during the \textit{gameplay} phase} and includes factual information (i.e., historical graphs and the current state of supply chain parameters), without introducing any form of bias into the players' performance. \newtext{After performance review, the Boss NPC would deliver the thought bubble prompt (see~\includegraphics[scale=0.75, trim=0 0.08cm 0 0]{figures/circle4.png} in Figure~\ref{fig:thoughtbubbles}).} \newtext{In terms of aesthetics, we did not design a thought bubble in literal terms. Instead, we chose to request players' thoughts via NPC dialogue as it felt natural to request this prompt as part of the meeting scenario, where the results are discussed, helping keep players in the context of the game.} \oldtext{Players experience this meeting scene once every four weeks (8 times total) throughout the entire game.}

\oldtext{We chose a four-week interval because, in doing so, we could frame the experience as a monthly meeting scene which felt more natural and resulted in the minimum distraction of players from their decision-making task. Moreover, our supply chain experiment includes a lead time of two weeks for ordering decisions (one week for orders to be processed by the manufacturer and one week for players to receive shipments). By considering a four-week interval, we allow players to have sufficient experience and observe the short-term outcome of their decisions before each thought bubble.}

To evoke players to reflect on their experience, we prompted them by asking, ``How do you think we are doing Kate?'' (see~\newtext{\includegraphics[scale=0.75, trim=0 0.08cm 0 0]{figures/circle3.png} in} Figure~\ref{fig:thoughtbubbles}) \newtext{and allowed them to provide open-ended input (see~\includegraphics[scale=0.75, trim=0 0.08cm 0 0]{figures/circle5.png} in Figure~\ref{fig:thoughtbubbles})}. We chose an open-ended question and response format to mitigate framing bias~\cite{tversky1985framing} and help players to articulate their thought process by engaging associative memory~\cite{memon2013enhanced}. The analyses we present in this paper center around players' responses to this question. Finally, \newtext{players experienced the meeting scene once every four weeks (8 times total) throughout the entire game (see~\includegraphics[scale=0.75, trim=0 0.08cm 0 0]{figures/circle_meeting.png} in Figure~\ref{fig:thoughtbubbles}), allowing us to collect data on their thought process over time. We chose a four-week interval because, in doing so, we could frame the experience as a monthly meeting scene which felt more natural and resulted in the minimum distraction of players from their decision-making task. Moreover, our supply chain experiment includes a lead time of two weeks for ordering decisions (one week for orders to be processed by the manufacturer and one week for players to receive shipments). By considering a four-week interval, we allow players to have sufficient experience and observe the short-term outcome of their decisions before each thought bubble.}\oldtext{in terms of aesthetics, we did not design a thought bubble in literal terms. Instead, we chose to request players' thoughts via NPC dialogue as it felt natural to request this prompt as part of the meeting scenario, where the results are discussed.} A video preview of the gamette and the thought bubbles is available in the OSF repository \newtextfinal{(\textcolor{cyan}{\url{https://osf.io/btfzx/?view_only=8211d2334d5440a0b75ae947811cb845}})}\oldtextfinal{(\textcolor{cyan}{\url{https://osf.io/btfzx/?view_only=d02d2f0447d945b1931d9d3a5fa953c6}})}.

\section{Study 1}\label{study1}
\begin{figure}[ht]
  \centering
  \includegraphics[width=0.5\columnwidth]{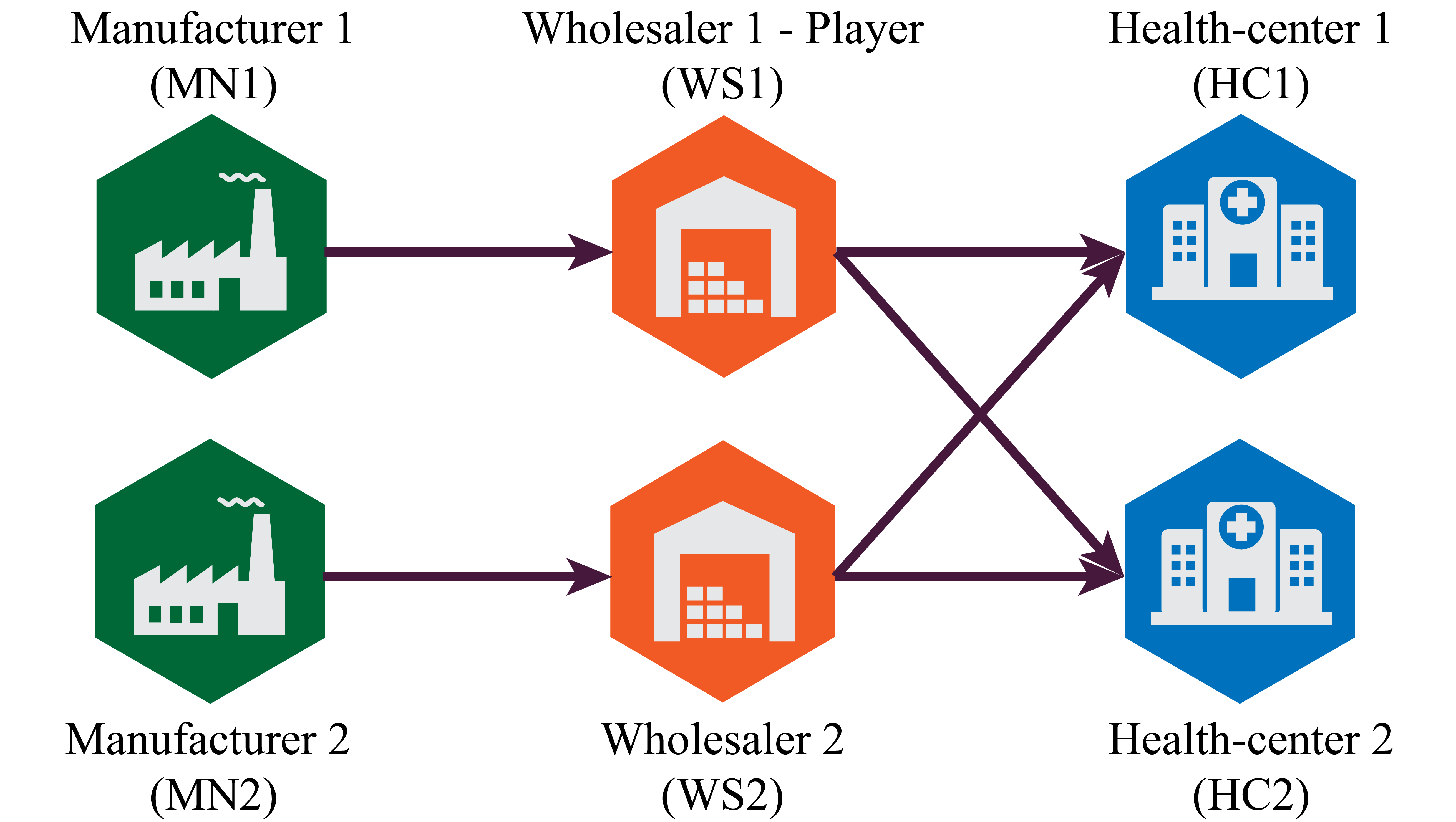}
  \caption{Supply chain network structure that included two manufacturers, two wholesalers and two health centers. Players play the role of Wholesaler 1 using the gamette.}
  \Description{The structure of supply chain network.}
  \label{fig:supply_chain_network}
  \Description[Supply chain network structure]{Structure of the supply chain network, including two manufacturers, two wholesalers, and two health centers. Players play the wholesaler role in the gamette.}
\end{figure}

The motivation behind Study 1 is to understand the role of human behavior in pharmaceutical supply chains experiencing drug shortages by studying the mental model development of decision makers. Specifically, we aim to elicit---through the thought bubbles---an individual decision maker's mental model and how this changes over time as they interact with the environment. 
To this end, we simulated a pharmaceutical supply chain network using the Flow Simulator and used a gamette to immerse human participants into a specific role within this network. We considered a supply chain network that includes two manufacturers, two distributors, and two health centers. Figure~\ref{fig:supply_chain_network} illustrates the network structure and the direction of shipments between each entity in this supply chain. We use this network because it permits the analysis of the supply chain network in conjunction with the behavioral dynamics that ensue from having multiple agents within each echelon~\cite{doroudi2020effects}.

\subsection{Methods}
\subsubsection{Hypotheses}
Drug shortages can occur for different reasons. However, most shortages can be traced back to supply chain disruptions~\cite{tucker2020incentivizing}. Many researchers have attempted to mitigate the impact of drug shortages through increasing supply chain resilience~\cite{tucker2020drug}, implementing decision support systems~\cite{chihaoui2019decision}, and developing optimal inventory management and ordering policies~\cite{azghandi2018minimization}. Another key, and often understudied, element is the behavior of human decision-makers, which can prolong or aggravate the effect of shortages~\cite{doroudi2020effects}. Such behaviors are exemplified in \newtextfinal{our} previous research \newtextfinal{where we} show\newtextfinal{ed}\oldtextfinal{ing that} people tend to deviate from optimal order suggestions~\cite{mohaddesi_introducing_2020}, and that these behaviors are attributed to hoarding and panic buying when facing shortages~\cite{mohaddesi_trust_2022}. This is due to people's tendency to ignore the temporal dynamics of the system when forming mental models of the supply chains~\cite{fu2006learning}. Additionally, different disruption locations can create different local dynamics for each role within the supply chain~\cite{rong2008impact,sarkar2016demonstrating}. These local dynamics present unique experiences for players, ultimately affecting their mental models and influencing their inventory management decisions, which contribute to the propagation of shortage effects throughout the rest of the supply chain. So our aim is to explore how players' mental models are affected when they are in different positions relative to the supply chain disruption, and hypothesize that: \\

H1. \textit{People show differences in their mental model development of a disrupted supply chain, depending on the disruption location.}\\

Previous research on drug shortage management also points to insufficient information sharing between stakeholders in pharmaceutical supply chains as a source for disruption propagation~\cite{yang2016current}, and that many of the costs associated with managing shortages in such supply chains can be mitigated by increasing collaboration and sharing information~\cite{pauwels2015insights}. Hence, information sharing is considered as one of the essential strategies to improve the resilience of supply chains~\cite{iyengar2016medicine}.
Information that is shared between supply chain stakeholders adds to the agents' management experiences in shaping their mental models of the surrounding supply chain dynamics. Thus, it is important to investigate the influence of information sharing on\oldtext{the formation of} mental model \newtext{development} and the behavior of decision-makers. We expect that sharing various levels of information results in the development of different mental models. Hence:\\

H2. \textit{People exhibit differences in mental model development of a disrupted supply chain, depending on the level of information being shared with them.}\\

\oldtext{We test these hypotheses by analyzing the frequency of qualitative codes, indicative of players' perceptions, comprehension, and projections (discussed in Section
that arise in players' responses to the thought bubbles prompt. 
}

\subsubsection{Experimental Design}\label{study1_experiment_design}
The experimental setting for our study is summarized in Table~\ref{tbl:experimental_design_study1}. Players in all conditions play the role of Wholesaler 1 (WS1; see Figure 3), and all other supply chain roles are assigned to agents that the Flow Simulator controls. The simulation agents make decisions based on order-up-to-level policy. According to this policy, each agent orders or produces enough product to bring their inventory position to a predefined level based on a periodic review policy with zero fixed costs~\cite{snyder2011fundamentals}. 
We considered different scenarios for testing the effect of disruption location on players' mental models (H1), where disruption can happen in either Manufacturer 1 (MN1) or Manufacturer 2 (MN2). We created these disruptions in the form of a manufacturing shutdown which reduces the production capacity of the disrupted manufacturer by 95\%.

\begin{table*}[ht]
    \caption{Summary of the experiment settings for Study 1.}
    \label{tbl:experimental_design_study1}
    \centering
    \renewcommand{\arraystretch}{1.6}
    \resizebox{0.8\textwidth}{!}{
    \begin{tabular}{l >{\centering\arraybackslash} p{60pt} >{\centering\arraybackslash} p{60pt} >{\centering\arraybackslash} p{60pt} >{\centering\arraybackslash} p{60pt} >{\centering\arraybackslash} p{60pt} >{\centering\arraybackslash} p{60pt}}
    \toprule
    & \textbf{\renewcommand{\arraystretch}{1} Condition 1} \newline \textbf{\renewcommand{\arraystretch}{1} (MN1/No-Info)} & \textbf{\renewcommand{\arraystretch}{1} Condition 2} \newline \textbf{\renewcommand{\arraystretch}{1} (MN2/No-Info)} & \textbf{\renewcommand{\arraystretch}{1} Condition 3} \newline \textbf{\renewcommand{\arraystretch}{1} (MN1/Partial-Info)} & \textbf{\renewcommand{\arraystretch}{1} Condition 4} \newline \textbf{\renewcommand{\arraystretch}{1} (MN2/Partial-Info)} & \textbf{\renewcommand{\arraystretch}{1} Condition 5} \newline \textbf{\renewcommand{\arraystretch}{1} (MN1/Complete-Info)} & \textbf{\renewcommand{\arraystretch}{1} Condition 6} \newline \textbf{\renewcommand{\arraystretch}{1} (MN2/Complete-Info)} \\ 
    \cmidrule{1-7}
    Player Role  & WS1 & WS1 & WS1 & WS1 & WS1 & WS1 \\
    Disrupted Manufacturer & MN1 & MN2 & MN1 & MN2 & MN1 & MN2 \\
    Information Sharing & No & No & MN1 inventory & MN1 inventory & MN1 inventory + HCs behavior and delivery rates & MN1 inventory + HCs behavior and delivery rates \\ 
    No. of Participants & 17	& 19 & 20 & 21 & 18 & 20 \\
    \bottomrule
    \end{tabular}}
\end{table*}

We also designed the two health-center agents (HC1 and HC2) in the Flow Simulator with different ordering behaviors to generate different local dynamics for the WS1 role based on disruption location. The difference in health-center behavior is related to how each health-center agent splits orders between wholesalers. While both health-centers receive a constant demand, the HC1 agent splits its orders to its upstream agents by means of a trustworthiness measure, meaning that HC1 orders less from the upstream wholesaler that fails to deliver drugs consistently. HC2, on the other hand, splits its orders always equally regardless of its wholesalers' trustworthiness. This trust-based behavior of HC1 particularly affects players during the shortage period (weeks 32-36) but differently depending on disruption location. When MN1 is disrupted, in addition to incurring stockout cost for not satisfying demand, players will also experience a decrease in HC1's demand. When MN2 is disrupted, players experience an increase in HC1's demand. According to these dynamics, we considered different options for testing the effect of information sharing on mental model development (H2): (1) without; (2) with information sharing on MN1 inventory (Partial Info); and (3) with information sharing on MN1 inventory, health-centers' ordering behavior, and delivery rates to each health-center (Complete Info). Figure~\ref{fig:orderingScene} and Figure~\ref{fig:alloc_tutorial_scenes} illustrate our designs for different levels of information sharing.


\subsubsection{Participants}
We recruited 115 participants ($87$ males, $26$ females, and $2$ not stated). Participants were full-time students enrolled in undergraduate and graduate level courses of Logistics and Supply Chain Management \newtextfinal{at Northeastern University}\oldtextfinal{in one of the higher education institutions in North America}. The age range is 18 to 32 years (\textit{M}=22.72, \textit{SD}=2.01). 

\subsubsection{Incentive Design}\label{study1_incentive_design}
Previous research points to the importance of incentives in conducting experimental research~\cite{katok2018designing}. To motivate participants to engage with the task and perform well, we offered them a monetary incentive (\$50) which was gifted through a raffle. Players who performed better had a higher chance in the raffle. Each participant received one ticket for completing the game plus one ticket for every \$1000 in-game profit that they made, more than the average profit of all other players. 

\begin{figure}[ht]
    \centering
    \begin{subfigure}[t!]{.48\columnwidth}
        \centering
        \includegraphics[width=0.95\columnwidth]{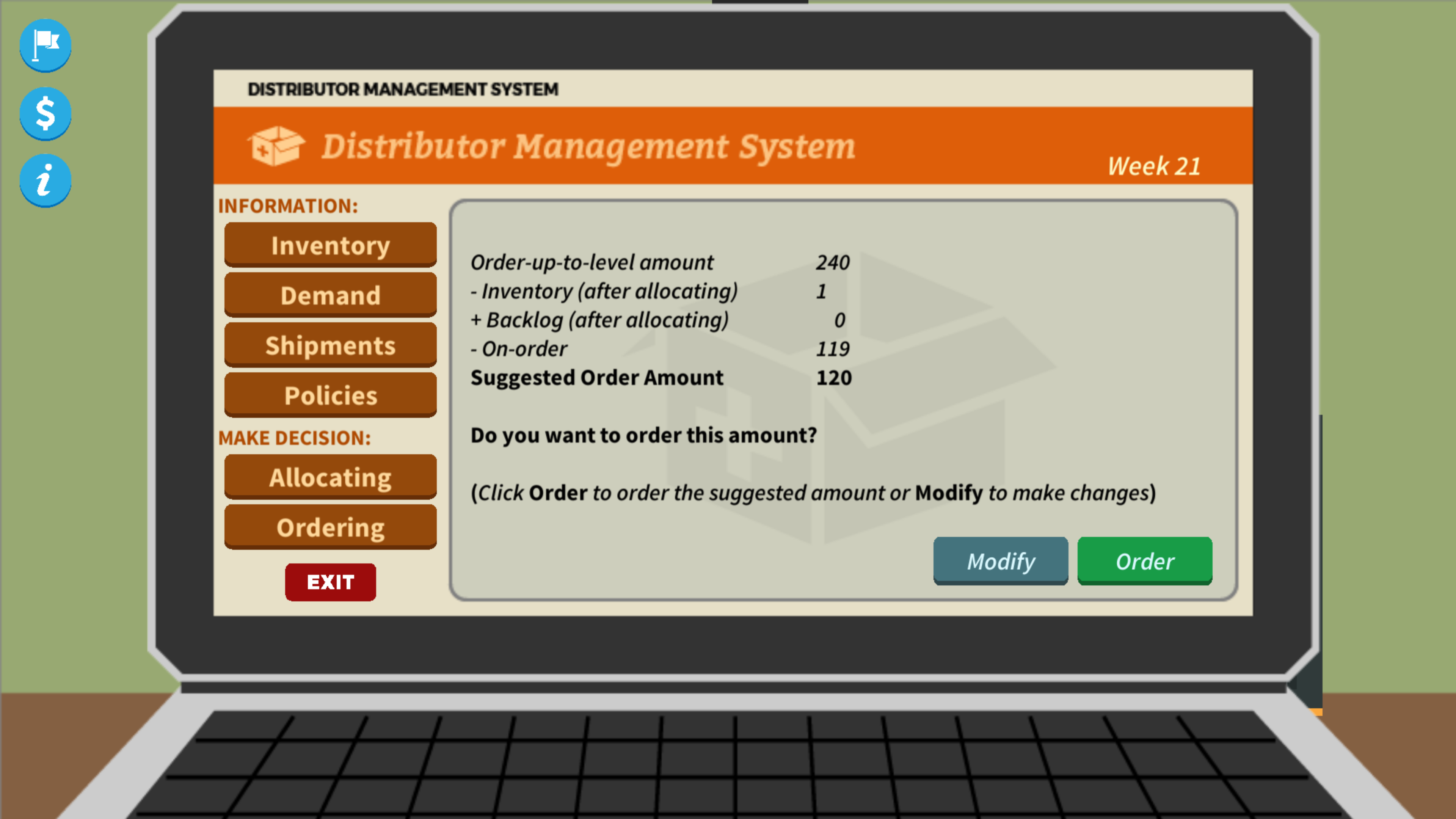}
        \caption{No information sharing on MN1 inventory }\label{fig:orderingScene1}
    \end{subfigure}
    \begin{subfigure}[t!]{.48\columnwidth}
        \centering
        \includegraphics[width=0.95\columnwidth]{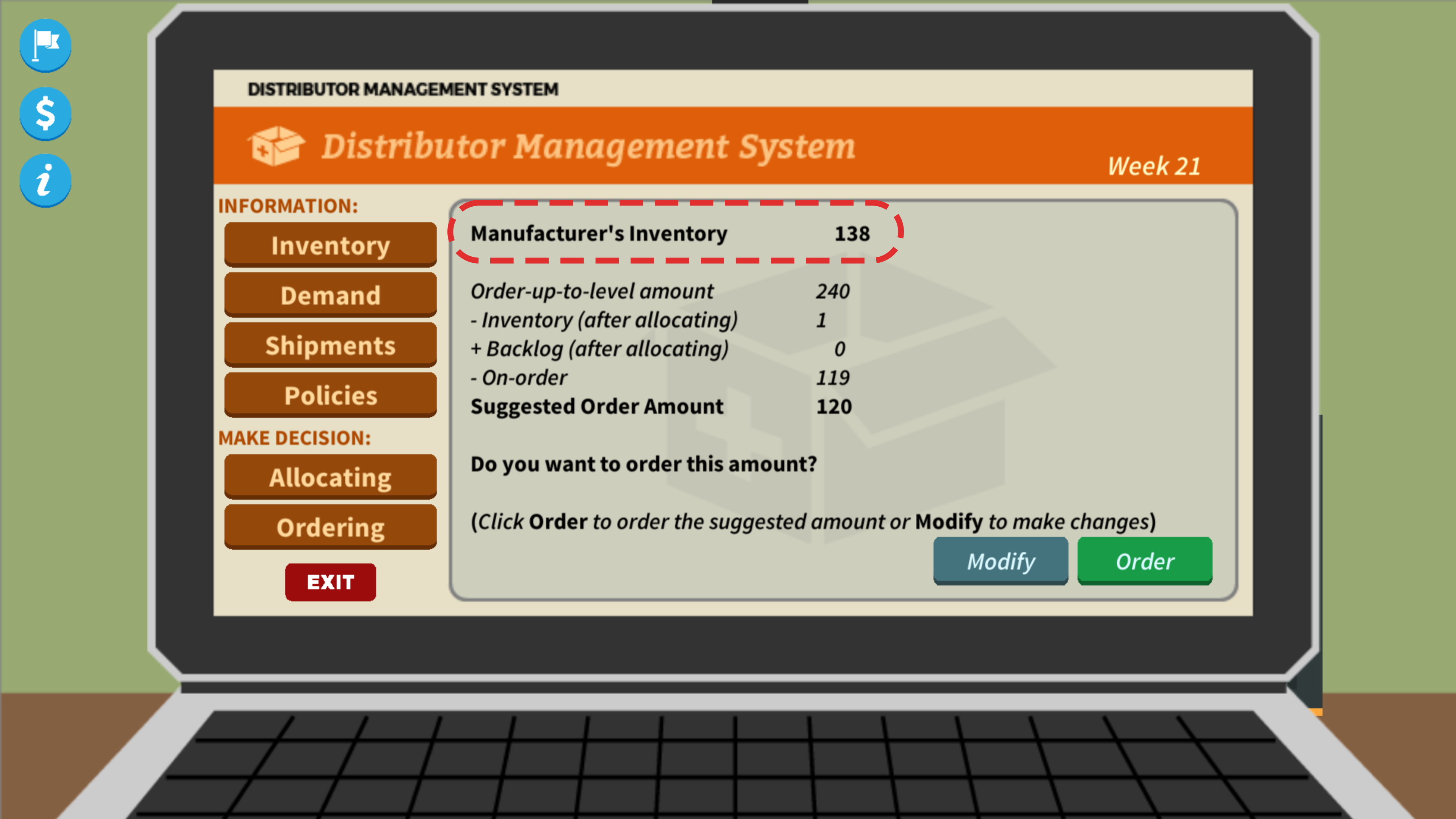}
        \caption{Sharing MN1 inventory}\label{fig:orderingScene2}
    \end{subfigure}
    \caption{The ordering scene as part of the decision task in (a) No-Info vs. (b) Partial and Complete Info conditions. Note that in (b)  ``Manufacturer's Inventory'' is mentioned at top while in (a) this information is not displayed. Players in both groups receive an order suggestion.}
    \label{fig:orderingScene}
    \Description[Difference between ordering scenes in conditions with and without information sharing in MN1 inventory]{Ordering scenes in different conditions look identical except for the manufacturer's inventory that is provided in the information sharing conditions.}
\end{figure}

\begin{figure}[ht]
    \centering
    \begin{subfigure}[t!]{.48\columnwidth}
        \centering
        \includegraphics[width=0.95\columnwidth]{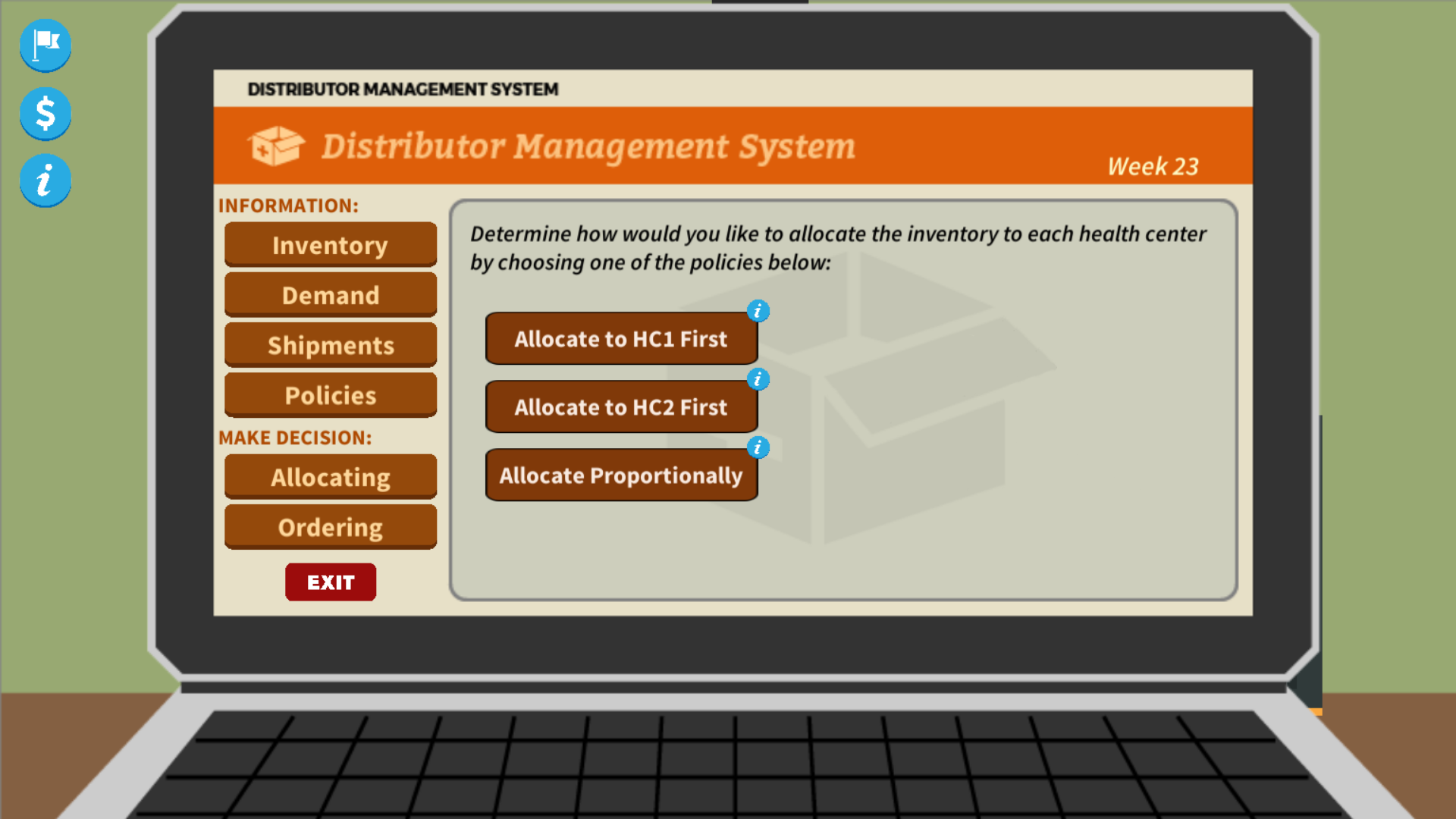}
        \caption{No information sharing on delivery rates to HCs}\label{fig:allocationscene1}
    \end{subfigure}
    \begin{subfigure}[t!]{.48\columnwidth}
        \centering
        \includegraphics[width=0.95\columnwidth]{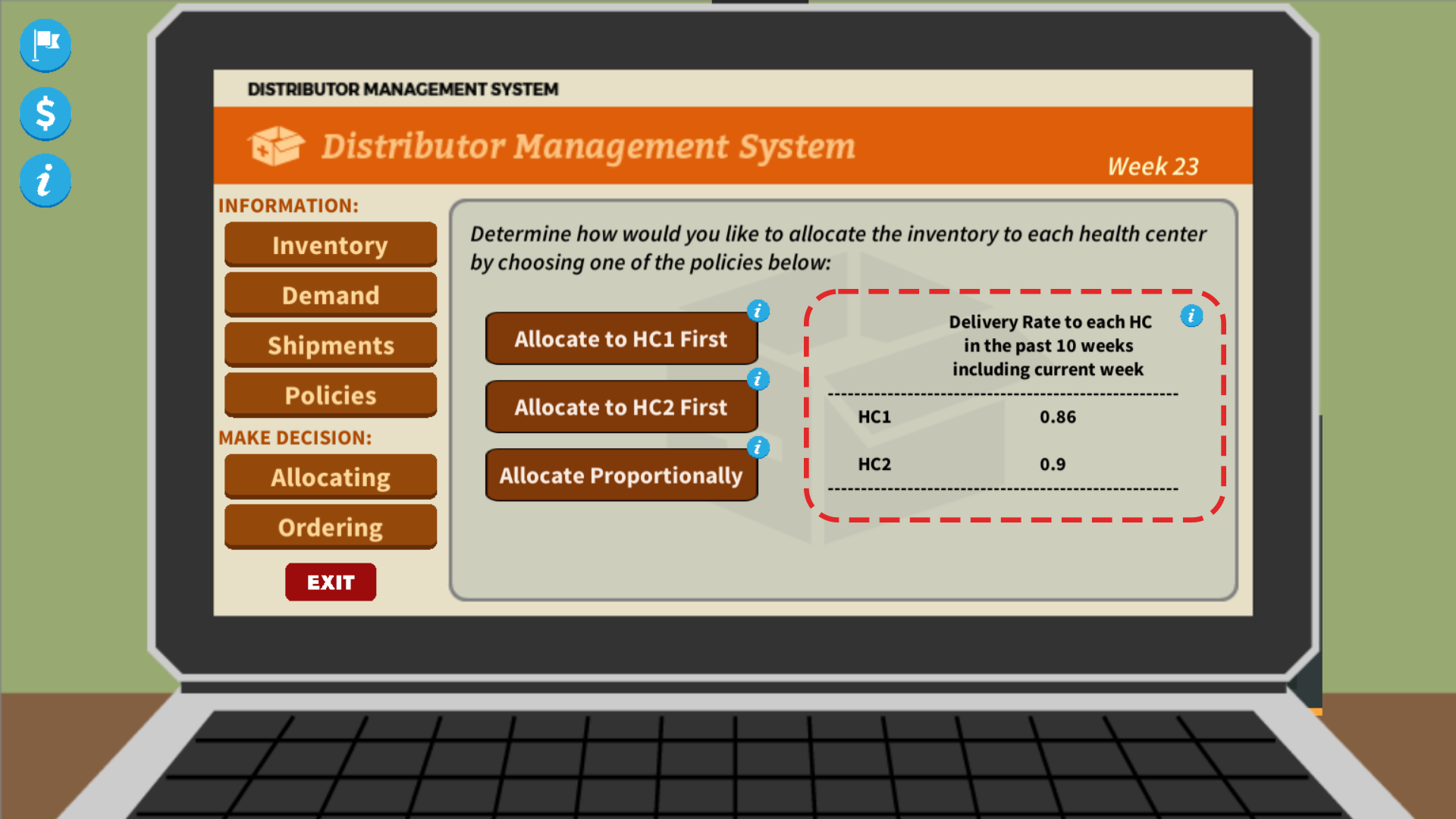}
        \caption{Sharing delivery rates to HCs}\label{fig:allocationscene2}   
    \end{subfigure}
    \vspace{.2cm}
    
    \begin{subfigure}[t!]{.48\columnwidth}
        \centering
        \includegraphics[width=0.95\columnwidth]{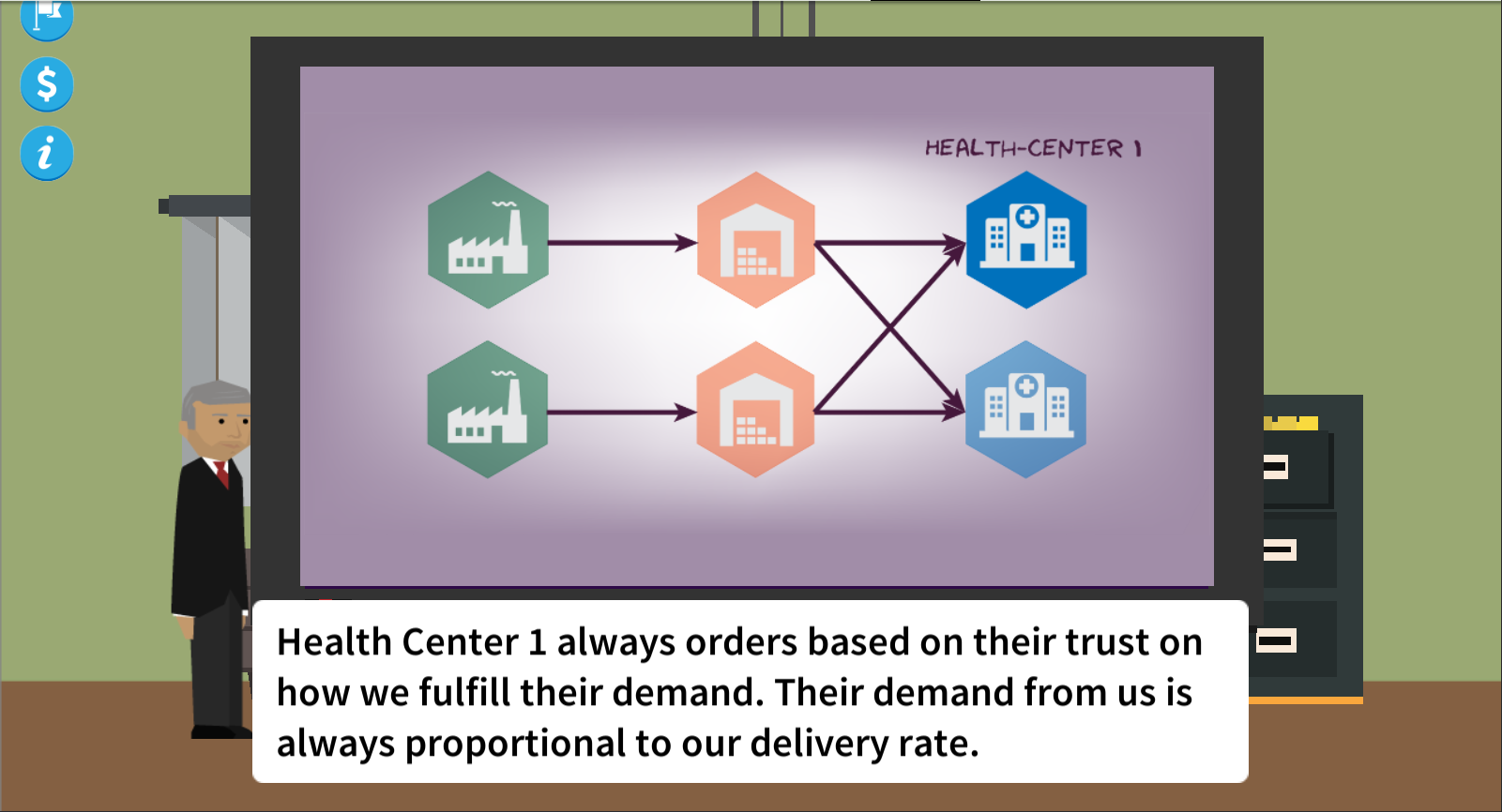}
        \caption{Information sharing on HC1 behavior}\label{fig:tutorialscene1}
    \end{subfigure}
    \begin{subfigure}[t!]{.48\columnwidth}
        \centering
        \includegraphics[width=0.95\columnwidth]{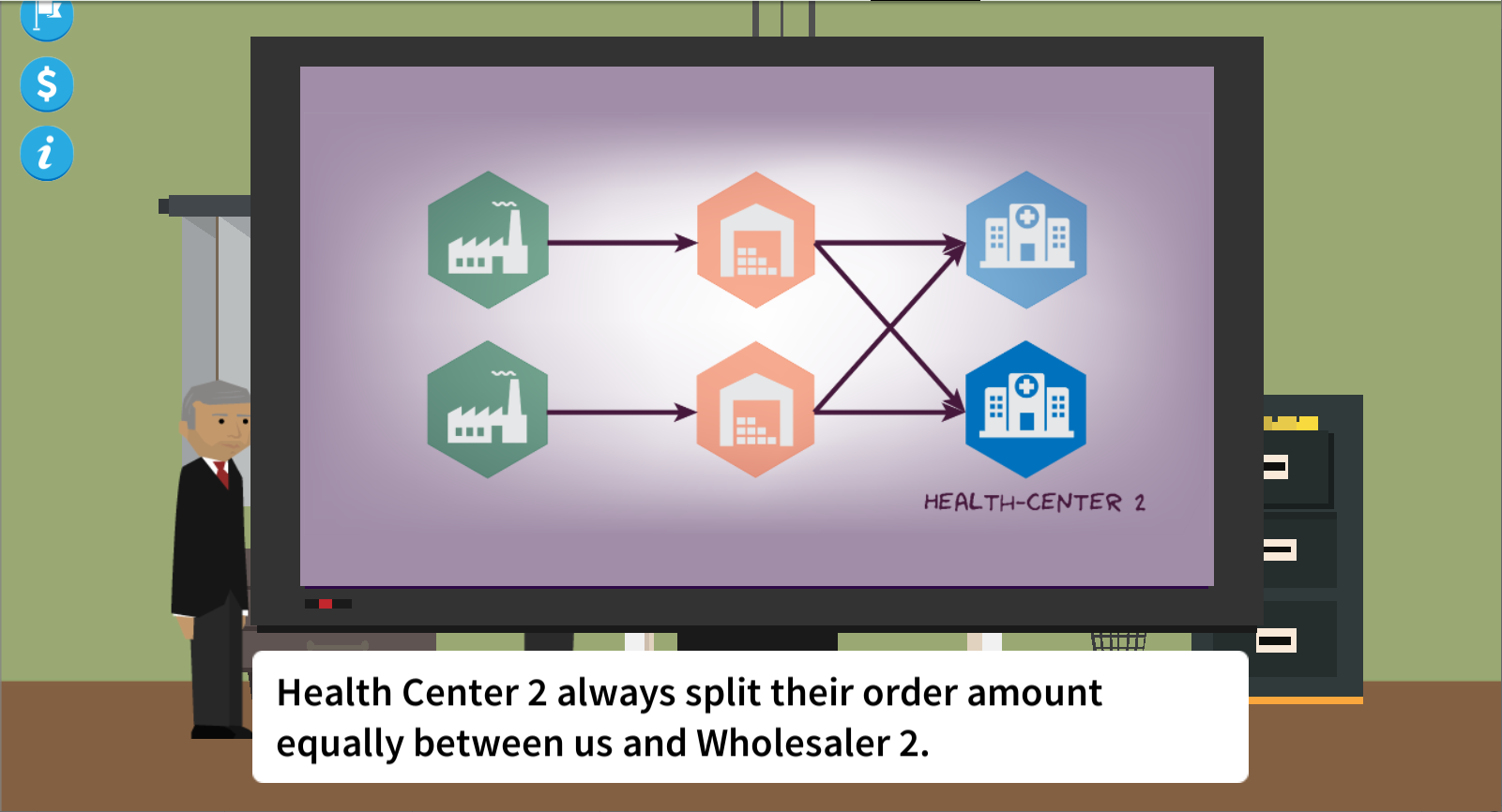}
        \caption{Information sharing on HC2 behavior}\label{fig:tutorialscene2}   
    \end{subfigure}
    \caption{The allocating scene (as part of the decision task) and tutorial scenes in (a) No-Info and Partial Info vs. (b-d) Complete Info conditions. Note that in (b)  ``Delivery Rate to each HC'' is mentioned in the allocating scene while in (a) this information is not displayed. (c) and (d) are also not presented to players in No-Info and Partial Info conditions.}
    \label{fig:alloc_tutorial_scenes}
    \Description[Difference between allocating and tutorial scenes in different conditions]{Allocating scenes in different conditions look identical except for the Delivery Rate to each HC that is provided in the complete information sharing conditions. In addition, the ordering behavior of health-centers are only displayed to players in Complete Info conditions during the tutorial phase.}
\end{figure}

\subsubsection{Procedure}\label{study1_procedures}
All participants first visited the study website, where they were formally briefed about the experiment and its purpose. By starting the game, each participant was randomly assigned to one of the six conditions. Table~\ref{tbl:experimental_design_study1} shows the number of participants assigned to each condition. The gamette in all conditions looked the same in all aspects except for the ordering scene, the allocating scene, and part of the tutorial that provided information on health-centers' ordering behavior. 
Participants played the role of a character named Kate who was hired as a supply chain director in a wholesaler company. At the beginning of the game, an NPC (Kate's boss) expresses that the game's goal is maximizing the company's profit by minimizing the inventory and stockout costs and maximizing sales revenue. The NPC also informs players, through dialogue, about the sales revenue and cost breakdown\oldtext{ (\$1 cost for each unit of inventory, \$10 cost for each unit of stockout, and \$5 revenue for each unit of sales)}, and the lead time of two weeks for orders (one week for orders to be processed by the manufacturer and one week for players to receive shipments).

Each participant first played four weeks of tutorial to familiarize themselves with the game (i.e., how to gather information) and was given instructions about ordering from the manufacturer and allocating to health-centers. Participants were asked to make an ordering decision at each period, but only make allocating decisions when their inventory level was lower than their total demand. In cases where they had enough inventory, the game automatically allocated drugs to each health-center. After finishing the tutorial, all sales and cost data was reset, and they played the game for 35 game weeks, starting at Week 21 of the simulation. After placing the order on Week 24, rather than moving to Week 25, the game transited players to the monthly meeting scene (Figure~\ref{fig:thoughtbubbles}) where they were queried via the thought bubble after receiving information on their performance via NPC dialogue. Players transited to this meeting scene every four weeks, with Week 52 being their last meeting. The disruption started on Week 28 and ended on Week 33. We framed the disruption as a manufacturing shutdown due to COVID-19.

At each period, participants received a shipment from their upstream manufacturer and could review the inventory, demand, and backlog information. Next, if they had limited inventory compared to demand, they were asked to select one of the presented allocation policies: (1) allocate to HC1 first, (2) allocate to HC2 first, or (3) allocate proportionally to each health center's demand. Finally, they received an order suggestion according to the order-up-to-level policy for making an ordering decision. They had the option to order the suggested amount or modify it. The gamette sends player decisions to the Flow Simulator, which moves the simulation to the next period and sends back the updated parameters for the next round to the gamette. After playing for 35 weeks, 
through dialogue, an NPC debriefed participants about the study, experimental setting, and their performance.

\subsubsection{Data Analysis}\label{data_analysis}
To analyze players' responses to the thought bubbles prompt (``How do you think we are doing Kate?''), we first applied Initial Coding~\cite{saldana2021coding}, where we qualitatively coded the open-response comments from players. We followed an inductive approach in our coding practice, where we used a combination of in vivo and constructed codes. Three researchers coded all players' comments across the eight prompts, leading to 94 initial codes. In further making sense of this data, we considered how these initial codes can lend insight into mental models by looking into existing theories and frameworks. We \newtextfinal{leveraged}\oldtextfinal{found that} the Situation Awareness (SA) model introduced by Endsley~\cite{endsley1995toward} (see Section~\ref{RW:Dynamic_Decision_Making}) \oldtextfinal{was helpful}to categorize our initial codes. 
We used three levels of SA to determine whether our generated codes reflect players' perception of the supply chain (in the game), their comprehension of the state relating to the game's goals, or their ability to project the future state of the supply chain. 

After assigning each code to one of the three SA levels, we determined what aspect of the supply chain (i.e., topic) is described in players' responses (e.g., profit, cost, inventory, demand, order, general, etc.) and how it is described through either a verb or adjective (i.e., description/action). The descriptions represented the rate of change (e.g., increase, decrease, constant), connotation (e.g., positive, negative, neutral) or specific actions (e.g., over-order, improve, anticipate, etc.) related to the supply chain topic. For example, we determined the comment ``inventory cost goes high'' to reflect a player's \textit{perception} (SA level) of an \textit{increase} (description) in \textit{inventory cost} (topic). The process took place by examining the codes through several group meetings. We did not seek to calculate an inter-rater reliability (IRR) as our generated codes were the process and not the product~\cite{mcdonald2019reliability}. We use the codes as a passage to get to the aspects of situation awareness and, from there, to get insight into mental model \newtext{development}\oldtext{formation}. Table~\ref{tbl:mental_models} summarizes the result of our qualitative analysis.

\oldtext{In our analysis, we realized that most players' comments reflect the comprehension aspect of SA. These comments are also primarily focused on a general description in response to the thought bubble prompt, with only a few words with a positive, negative or neutral connotation, which are exemplified in comments such as ``good'', ``great'', ``bad'', ``pretty bad'', ``okay'', or ``fair''. While one can be skeptical of a single word representing an individual's mental models, we argue that in the context of the situation models, these words represent the verbalization of a mental process about how players can or cannot make sense of their environment in general. We provide more details on the insights that such brief comments can provide in testing our hypotheses in Section
}

To understand players' mental model development, we performed quantitative analyses in three steps: (1) counting the number of codes representing each aspect of SA in player comments and performing hypothesis testing to investigate associations; (2) studying patterns by examining the counts of codes temporally and across the eight measurements; and (3) explaining the patterns and associations by examining the topics/descriptions of SA aspects as represented in players' comments. All comments, codes, and generated categories (i.e., topics and descriptions/actions) through our qualitative analysis are available in the OSF repository \newtextfinal{(\textcolor{cyan}{\url{https://osf.io/btfzx/?view_only=8211d2334d5440a0b75ae947811cb845}})}\oldtextfinal{(\textcolor{cyan}{\url{https://osf.io/btfzx/?view_only=d02d2f0447d945b1931d9d3a5fa953c6}})}. In Section~\ref{study1_results} and in Table~\ref{tbl:mental_models}, we refer to participant quotes as ``(Player ID, gender, age).'' For example, a female player aged 23 with player ID 85 who participated in Study 1 would be displayed as ``(PL1-85, female, 23).''

\begin{table*}[ht]
\caption{Aspects of mental models identified by applying SA to qualitative responses of players to the thought bubbles.}
    \label{tbl:mental_models}
    \centering
    \resizebox{.75\textwidth}{!}{
    \begin{tabular}{lp{3cm}p{3cm}p{5cm}}
    \toprule
    SA Level & Topic & Description/Action & Example \\
    \cmidrule{1-4}
    Perception      & inventory cost, backlog cost, costs, profit, inventory, demand, backlog, order & increase, decrease, consistent, zero , over-order, under-order & \textit{``there is uncertainity [sic] in supply, thats [sic] why our \textbf{inventory cost goes high}, we need to talk with supplier.''}--(PL1-18, male, 23) \newline \textit{``I think we did pretty well. \textbf{The profit was constant and the cost was minimal}. All orders were satisfied.''}--(PL1-112, female, 23)" \\ \\
    Comprehension   & general, inventory, demand, backlog, supply line, order & positive, negative, neutral, uncertain & \textit{``\textbf{Awful}. Need to introspect and make the necessary changes.''}--(PL1-3, male, 24) \newline \textit{``We are doing \textbf{fantastic} and \textbf{meeting demands} regularly with \textbf{enough saftey [sic] stock in hand} to satisfy sudden hike in the demand.''}--(PL1-2, male, 24)" \\ \\
    Projection      & general, profit, inventory, demand, backlog, order, allocation & improve, anticipate problem/uncertainty, increase, decrease, constant, uncertain, proportionally, HC with higher delivery rate, HC2 & \textit{``We have more inventory since the manufacturer 2 is closed due to COVID-19 so \textbf{we have to prepare for the incresing [sic] of Healthcare 2 demand}.''}--(PL1-47, female, 25) \newline \textit{``Profit is going down and performing poorly due to high backorder costs. \textbf{I need to order more.}''}--(PL1-98, male, 23)" \\ 
    \bottomrule
    \end{tabular}}
\end{table*}

\subsection{Results}\label{study1_results}
Participants wrote on average six words per prompt (\textit{Mdn}=$3$, \textit{IQR}=$7$). The thought bubble prompt was left unanswered in 5.5\% of the instances across the eight measures, spreading across 19 participants. \newtext{The mean time players spent in the meeting scene was 55.1 seconds (\textit{SD}=69.7), which included the time they reviewed their performance and responded to the prompt.} We report these statistics to illustrate how thought bubbles are utilized. The following sections describe what we learned from players' responses in Study 1.

We used the outcome of our qualitative analysis and mental model development to compute the frequency of players' comments that reflected each aspect of SA depending on the disruption location (see Table~\ref{tab:contingencyTables}). Chi-square test of independence revealed significant association between disruption location and players' mental models, as captured by SA via thought bubbles (\newtext{${\chi}^2_{2, 1011}=12.047$, }$p=.002$)\newtext{, and indicated a small effect (Cramer's $V=.109$)~\cite{cohen2013statistical}}. More specifically, players who experienced a disruption in their own supplier (i.e., MN1) wrote significantly more comments reflecting their comprehension of the environment ($p<.001$). These comments mostly indicate general negative expressions, such as:

\begin{quote}
    \textit{``Awful. Need to introspect and make the necessary changes.''}--(PL1-3, male, 24) \\
    \textit{``very bad.''}--(PL1-66, female, NA) \\
    \textit{``We're doing terribly since our supplier is unable to provide more than 20 shipments per weeky [sic] order.''}--(PL1-74, male, 21) \\
    \textit{``We are not recieving [sic] orders, that's big problem.''}--(PL1-13, male, 24)
\end{quote}

Higher comprehension frequency in the comments of this group (i.e., MN1 disruption) does not necessarily reflect a better understanding of the environment and instead shows the level of uncertainty or inability to make sense of the environment towards goals. We also investigated the progression of mental models over time (see Figure~\ref{fig:study1_sa_over_time_players_and_mainpulation}-A2). We realized that most negative expressions of players with MN1 disruption are mainly observed at the end of the shortage period (i.e., Week 36). This can indicate that the generated dynamics by disruption in MN1 create more uncertainty for players to the extent that they primarily comprehend their environment as negative. Therefore, the source of disruption affects each player's mental model development relative to their comprehension.

\begin{table}[ht]
	\centering
	\caption{Contingency table of frequency of players' comments representing each aspect of Situation Awareness (SA), separately for implemented manipulations and across the eight thought bubbles prompts in Study 1.}
	\label{tab:contingencyTables}
	\resizebox{.75\columnwidth}{!}{
		\begin{tabular}{lrrrrr}
			\toprule
			\multicolumn{1}{c}{} & \multicolumn{1}{c}{} & \multicolumn{3}{c}{SA Level} & \multicolumn{1}{c}{} \\
			\cline{3-5}
			
			Manipulation &  & Perception & Comprehension & Projection & Total  \\
			\cmidrule[0.4pt]{1-6}
			Disrupted Location \\[.1cm]
			\quad MN1 ($n=55$) & Count & 69.0 & 358.0\rlap{\textsuperscript{$\ast\ast$}} & 38.0\rlap{\textsuperscript{$\ast$}} & 465.0  \\
			 & Expected & 79.1 & 334.3 & 51.5 & 465.0  \\
			\quad MN2 ($n=60$) & Count & 103.0 & 369.0\rlap{\textsuperscript{$\ast\ast$}} & 74.0\rlap{\textsuperscript{$\ast$}} & 546.0  \\
			 & Expected & 92.9 & 392.6 & 60.4 & 546.0  \\[.1cm]
			 \cmidrule[0.4pt]{1-6}

			Information Sharing \\[.1cm]
			\quad Complete ($n=38$) & Count & 60.0 & 244.0 & 34.0 & 338.0 \\
			 & Expected & 57.5 & 243.0 & 37.4 & 338.0 \\
			 
			\quad Partial ($n=41$) & Count & 80.0\rlap{\textsuperscript{$\ast$}} & 246.0\rlap{\textsuperscript{$\ast\ast$}} & 51.0 & 377.0 \\
			 & Expected & 64.13 & 271.1 & 41.7 & 377.0 \\
			
			\quad No-Info ($n=36$) & Count & 32.0\rlap{\textsuperscript{$\ast\ast$}} & 237.0\rlap{\textsuperscript{$\ast\ast$}} & 27.0 & 296.0 \\
			 & Expected & 50.3 & 212.8 & 32.8 & 296.0 \\[.1cm]
			 \cmidrule[0.4pt]{1-6}
			 
			Total & Count & 172.0 & 727.0 & 112.0 & 1011.0  \\
			 & Expected & 172.0 & 727.0 & 112.0 & 1011.0  \\
			
			\bottomrule
			\multicolumn{6}{m{12cm}}{\small Counts show the presence of each SA level in players' responses. \newtext{Expected represents the expected counts for the Chi-square test under the null hypothesis (i.e., no association between manipulations and SA levels.)}}\\
            \multicolumn{6}{m{12cm}}{\textsuperscript{$\ast$}\small Chi-square Post-hoc test shows significant association between experimental manipulations and SA level at $\alpha=0.05$ with Bonferroni correction.}\\
            \multicolumn{6}{m{12cm}}{\textsuperscript{$\ast\ast$}\small Chi-square Post-hoc test shows significant association between experimental manipulations and SA level at $\alpha=0.01$ with Bonferroni correction.}
		\end{tabular}
	}
\end{table}

			

Our analysis also showed that a disruption in MN1 is associated with significantly lower projection, as evidenced by players' comments ($p=.003$). Such discrepancy can be mainly attributed to the comments at the end of the shortage period (see Figure~\ref{fig:study1_sa_over_time_players_and_mainpulation}-A3), where comments from players with a MN1 disruption in Week 36 reflect significantly fewer projections ($p=.002$). As players with a MN2 disruption experience an increase in their demand, their projections represent topics such as ``anticipating uncertain demand'', ``anticipating demand to increase,'' or ``planning to over-order'':

\begin{quote}
    \textit{``Doing well so far. I expect it will be difficult once the demand becomes uncertain due to covid19.''}--(PL1-12, male, 25) \\
    \textit{``demand from HC1 is increasing and we have to supply accordingly.''}--(PL1-56, male, 24)
\end{quote}

Finally, the chi-square post-hoc test did not show a significant difference in the frequency of players' perceptions depending on disruption location when considering all thought bubble responses. However, we found that players with MN1 disruption represent significantly lower perception in their responses during Week 36 ($p=.005$) (see Figure~\ref{fig:study1_sa_over_time_players_and_mainpulation}-A1). Players experiencing a MN1 disruption seem to only perceive the ``increase in their backlog'', ``increase in backlog cost'', and ``decrease in profit'':

\begin{quote}
    \textit{``So bad, first I wanted to keep the demand of HC1, but then I found more demand increase more backlogs which cost more money.''}--(PL1-7, male, 22) \\
    \textit{``Net profit is negative because of backlog cost.''}--(PL1-113, male, NA)
\end{quote}

The disruption location affects the state of the supply chain for a given point in time such that players in different groups experience different dynamics. Our results show that these emergent dynamics caused by different disruption locations drive the \newtext{development}\oldtext{formation} of different mental models in how players comprehend the environment and project into the future in general, and how they perceive the system when facing a shortage---thus providing support for H1.

\subsubsection{Mental Model Development vs. Level of Information Sharing (H2)}
Chi-square test of independence also showed a significant association between the level of information and mental model development (\newtext{${\chi}^2_{4, 1011}=19.172$, }$p<.001$). \newtext{The effect size for this finding was small (Cramer’s $V=.097$).} The results of the chi-square post-hoc test (see Table~\ref{tab:contingencyTables}) did not reveal a significant association between complete information sharing and the development of mental models. However, our results indicated that the responses of players who only received information on their supplier inventory (i.e., partial information) represented significantly higher perception ($p=.003$) and significantly lower comprehension ($p<.001$).
As for the perception, players in the Partial-Info group seem to speak to almost all topics more frequently than players in No-Info group, as exemplified in:

\begin{quote}
    \textit{``We are in very high loss but may fulfil their requierments [sic] as manufacturing unit have more units readily available.''}--(PL1-44, male, 22) \\
    \textit{``we have done really well over the past 2-3 weeks. we've been able to keep up with demand while maintaining low costs. I will look to keep the inventory low for the time being.''}--(PL1-73, male, 21) \\
\end{quote}

\begin{figure*}[ht]
    \centering
    \includegraphics[width=.8\textwidth]{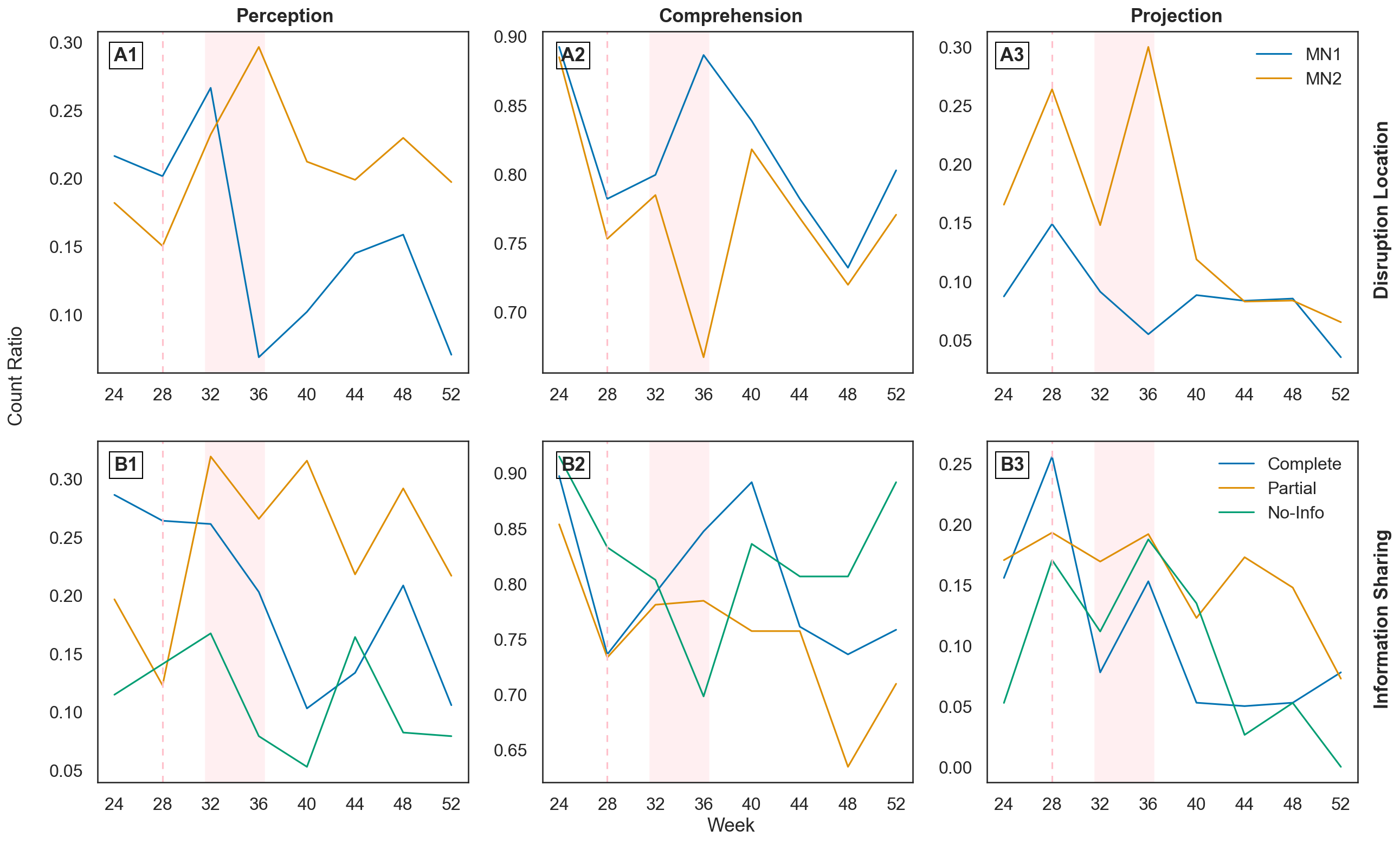}
    \caption{Progression of different aspects of situation model over time depending on disruption location and level of information sharing in Study 1. The vertical dashed line in each graph shows the time players were notified about the disruption. The highlighted area illustrates the shortage period due to manufacturing disruption. \newtext{For each experimental group, count ratio was calculated by dividing the number of comments that reflect each SA aspect by total number of players in that group.} 
    }\label{fig:study1_sa_over_time_players_and_mainpulation}
    \Description[lineplot showing average count of SA concepts in Study 1]{six lineplots showing the average count of SA concepts in Study 1 for different disruption locations and information sharing levels}
\end{figure*}

The higher perception of players in this group is also reflected in the average count of comments over time. Players in the Partial-Info group show an increase in the average number of comments related to perception when the shortage starts in Week 32 (see Figure~\ref{fig:study1_sa_over_time_players_and_mainpulation}-B1). On the other hand, the lower frequency of comments related to comprehension in the Partial-Info group is mainly reflective of general positive comments:

\begin{quote}
    \textit{``the past several weeks have been good, although manufacturer 2's accident in Puerto Rico may change things up...''}--(PL1-73, male, 21) \\
    \textit{``Its [sic] look market is working good but still we keep high inventory due to uncertainity''}--(PL1-44, male, 22)
\end{quote}

These players seem to be more cautious in interpreting the environment as positive because of having access to partial information. Finally, chi-square post-hoc test indicated that not having access to information is associated with significantly lower perception ($p<.001$) and significantly higher comprehension ($p<.001$). Higher comprehension of players in the No-Info group reflects codes ``general-positive'' and ``general-negative''. As the results indicate, not having access to information makes it challenging for players to make sense of the environment. On the other hand, partial information directs players' attention to perceive more, and complete information has no effect, perhaps due to information overload~\cite{endsley1995toward}. 
These results show partial evidence for H2, where information sharing on supplier inventory affects\oldtext{the formation of} mental model \newtext{development} of disrupted supply chains. 

\newtext{\subsection{Summary}\label{study1_summary}
Study 1 served as the foundation for exploring the use of thought bubbles for eliciting mental model development. We obtained a codebook (Table~\ref{tbl:mental_models}) by examining players' comments through qualitative analysis and grounding it in SA theory. This codebook represented the aspect of players' situation models in our decision-making task, which has shown to be directly related to mental models~\cite{endsley2015situation}. It is crucial to test the reliability of this approach for eliciting mental model concepts. Hence, we will use our codebook to qualitatively code responses to the thought bubbles prompt for new players in Study 2. Study 1 also demonstrated how experimental manipulations affect mental model development over time. More specifically,
\begin{itemize}
    \item a disruption in players' suppliers affects their mental model development by directing them to comprehend the environment as unfavorable and projecting less about how the future unfolds.
    \item receiving information on supplier inventory directs players' attention to key features of the environment (i.e., more perception) and causes them to be more cautious in how they interpret the situation.
    \item sharing no information causes players to perceive the elements of the system less and face difficulty in making sense of the environment.
\end{itemize}
In examining mental model development in Study 1, however, we did not consider the observed behavior of players. Therefore, in the subsequent study, we examine if the observed decisions of players---as reflected in their behavioral profiles---are the outcome of different mental model development processes. 
}

\section{Study 2}\label{study2}
To get insight into how mental models drive human decisions, we need to investigate another key aspect influencing mental model development: individual differences. Prior research provides evidence of different decision patterns among supply chain decision makers~\cite{mohaddesi_trust_2022, sun2016modeling}. Therefore, we conducted Study 2 to \newtext{explain the cognitive aspect of decisions made by players with different behavioral profiles, while} test\newtext{ing} the reliability of our generated codebook in \newtext{Study 1.}\oldtext{studying mental model development across different behavioral profiles.} We also made some changes to the experimental setting of Study 1. First, since we did not find evidence for the effect of complete information sharing, we decided to drop that as a manipulation for Study 2. Second, we decided to limit the disruption location to MN1 to isolate the generated dynamics for easier comparison of mental model developments across players with distinct behavioral profiles.

\subsection{Methods}

\subsubsection{Hypotheses}
Individuals show different behaviors which might be affected by various factors~\cite{appelt2011decision}. In the context of dynamic decision-making, one can attribute these behavioral differences to the changes in mental model development~\cite{salvendy2012handbook}. The literature around SA also points to individual characteristics affecting SA~\cite{endsley2015situation, endsley2000situation_book}, which in turn can affect mental model \newtext{development}\oldtext{formation}. Therefore, we expect to see associations between observable behavioral profiles and mental model \newtext{development}\oldtext{formation} as captured via thought bubbles. Hence, we hypothesize: \\

H3. \textit{People show differences in mental model development of a disrupted supply chain dynamics, depending on their behavioral profile.}\\

We retrieve behavioral profiles following the methods \newtextfinal{we} proposed \newtextfinal{in our previous work}\oldtextfinal{by Mohaddesi et al.}~\cite{mohaddesi_trust_2022} (see~\ref{study_2_data_analysis}) to test this hypothesis. Furthermore, in continuation of Study 1, we investigate how information sharing affects mental model formation for different behavioral profiles. While we did not find significant association between complete information sharing and SA levels, we found that access to partial information (i.e., supplier's inventory level) is associated with significantly higher perception and lower comprehension. In \newtextfinal{our previous}\oldtextfinal{their} study, \oldtextfinal{Mohaddesi et al.}\newtextfinal{we} provided empirical evidence on how information sharing affects the behavior of players with different decision patterns~\cite{mohaddesi_trust_2022}. Consequently, we expect the effect of information sharing on observable actions to be reflected in players' mental model development as captured via thought bubbles. Hence:\\

H4. \textit{People with different behavioral profiles show differences in mental model development of a disrupted supply chain, depending on the level of information being shared with them.}\\ 

We test these hypotheses using the same approach as in Study 1. 

\subsubsection{Experimental Design}\label{study2_experimental_design}
Table~\ref{tbl:experimental_design_study_2} summarizes the experiment settings for Study 2, where we considered a disruption in MN1 and two options for information sharing: (1) without, and (2) with information sharing on MN1 inventory. This dataset contains players' responses to the thought bubbles prompt collected following the same procedures described in Section~\ref{study1_procedures}. All other aspects, including supply chain roles and disruption settings, are identical to Study 1 (see Section~\ref{study1_experiment_design}).

\subsubsection{Participants}\label{study2_participants}
Through Prolific~\footnote{\url{www.prolific.co}}, we recruited 135 (57 males, 74 females, 2 non-binary, 2 not stated) online participants. We only allowed participation for working professionals who had an undergraduate degree or higher and reported English as their first language or were fluent in it. Each participant spent, on average, 58 minutes (\textit{SD}=19.35) playing and received a \$7.5 reward for their participation. The age range is 21 to 71 years (\textit{M}=34.23, \textit{SD}=9.86). We provided participants with the same incentive as in Study 1 (see Section~\ref{study1_incentive_design}) and followed the same procedures (see Section~\ref{study1_procedures}).



\begin{table*}[ht]
    \caption{Summary of the experiment settings in Study 2.}
    \label{tbl:experimental_design_study_2}
    \centering
    \renewcommand{\arraystretch}{1.6}
    \resizebox{0.8\textwidth}{!}{
    \begin{tabular}{lccccccc}
    \toprule
    & \multirow{2}{1cm}{\centering Player Role} & \multirow{2}{2cm}{\centering Disrupted Manufacturer} & \multirow{2}{2cm}{\centering Information Sharing} & \multicolumn{4}{c}{No. of Participants} \\
    \cline{5-8} 
    & & & & Hoarders & Reactors & Followers & Total \\ 
    \hline 
    Condition 1 (No-Info) & WS1 & MN1 & No & 31 & 21 & 9 & 61\\
    Condition 2 (Info) & WS1 & MN1 & Yes & 25 & 27 & 8 & 60\\
    No. of Participants & & & & 56 & 48 & 17 & 121\\ 
    \bottomrule
\end{tabular}}
\end{table*}

\subsubsection{Data Analysis}\label{study_2_data_analysis}
The three researchers involved in data analysis of Study 1 qualitatively coded players' open-response comments (a total of 1093 comments) by using the previously generated categories as a codebook (see Table~\ref{tbl:mental_models}). Each researcher used a tuple in the form \textit{<SA Level, Topic, Discription/Action>} to code each comment within the dataset. In case a player's comment reflected multiple topics or SA aspects simultaneously, we used multiple tuples to represent them. Our goal was to evaluate whether data are being interpreted in the same way relative to the previous analysis in Study 1. Therefore, per the recommendations of McDonald et al.~\cite{mcdonald2019reliability}, we attempted to measure inter-rater reliability (IRR) by calculating Fleiss' kappa ($\kappa$)~\cite{fleiss1971measuring}. After the first round of coding, we obtained $\kappa$=$0.75$. While this showed a relatively high agreement rate, we examined the codes once more to address sources of disagreement. After our second attempt, we obtained $\kappa$=$0.84$, which we considered sufficient for our hypothesis testing. 

The steps for our quantitative analysis were similar to those mentioned in Study 1 (see~\ref{data_analysis}). However, to decide which set of codes to use for the rest of our analysis, we considered two heuristics: (1) using the majority vote (i.e., codes in which two out of three coders agreed upon), and (2) using codes from the master coder (i.e., first author) in addition to the majority vote. During our analysis, we realized that both heuristics show similar results. Therefore, for simplicity, we chose to move forward with the first heuristic.
Players' comments and qualitative codes for Study 2 can be accessed in the OSF repository \newtextfinal{(\textcolor{cyan}{\url{https://osf.io/btfzx/?view_only=8211d2334d5440a0b75ae947811cb845}})}\oldtextfinal{(\textcolor{cyan}{\url{https://osf.io/btfzx/?view_only=d02d2f0447d945b1931d9d3a5fa953c6}})}. 

To obtain the behavioral profile of players, we followed the methods \newtextfinal{from our prior study}\oldtextfinal{proposed by Mohaddesi et al.}~\cite{mohaddesi_trust_2022}. We first filtered outlier players \newtext{(\textit{n}=$14$)} and then used a Hidden Markov Model (HMM) to infer players' response modes in their deviation from order suggestions (see Figure~\ref{fig:orderingScene} for an example of order suggestion). Finally, through cluster analysis of sequences of response modes, we found three behavioral profiles: (1) Hoarders who deviate from order suggestions and over-order more frequently, (2) Reactors who follow order suggestions but react to the disruption news and deviate from recommendation only after disruption, and (3) Followers who almost always follow order suggestions. For more detailed description of the methods for retrieving behavioral profiles we refer to~\cite{mohaddesi_trust_2022}. Table~\ref{tbl:experimental_design_study_2} shows the count of players characterized by each behavioral profile across the two conditions. Moving forward, we investigate the mental model development (as captured with thought bubbles) of these three profiles in testing our hypotheses.
\subsection{Results}
Similar to Study 1, we investigated how thought bubbles were utilized among Study 2 participants. 
On average, players wrote five words in each of their comments (\textit{Mdn}=$2.00$, \textit{IQR}=$5.25$). The thought bubble prompt was left unanswered in 3.8\% of the instances across the eight measures, which were spread across 12 players. \newtext{Players spent, on average, 56.4 seconds (\textit{SD}=47.3), where they reviewed their performance and responded to the prompt.}

\begin{table}[ht]
	\centering
	\caption{Contingency table of frequency of players' comments representing each aspect of Situation Awareness (SA), separately for implemented manipulations and across the eight thought bubbles prompts in Study 2.}
	\label{tab:contingencyTables_study2}
	\resizebox{.75\columnwidth}{!}{
		\begin{tabular}{clrrrrr}
			\toprule
			& \multicolumn{1}{c}{} & \multicolumn{1}{c}{} & \multicolumn{3}{c}{SA Level} & \multicolumn{1}{c}{} \\
			\cline{4-6}
			& Profile/Manipulation &  & Perception & Comprehension & Projection & Total \\
			\cmidrule[0.4pt]{1-7}
			\multirow{6}{*}{\rotatebox[origin=c]{90}{Behavior Profile}} & Hoarder & Count & 88.0 & 387.0\rlap{\textsuperscript{$\ast$}} & 56.0 & 531.0 \\
			 & & Expected & 75.5 & 404.8 & 50.6 & 531.0 \\
			& Reactor & Count & 60.0 & 333.0 & 42.0 & 435.0 \\
			 & & Expected & 61.8 & 331.6 & 41.5 & 435.0 \\
			& Follower & Count & 4.0\rlap{\textsuperscript{$\ast\ast$}} & 95.0\rlap{\textsuperscript{$\ast\ast$}} & 4.0 & 103.0 \\
			 & & Expected & 14.6 & 78.5 & 9.8 & 103.0 \\
			\cmidrule[0.4pt]{1-7}
			
			
			\multirow{17}{*}{\rotatebox[origin=c]{90}{Information Sharing per Behavior Profile}} & Hoarder \\ [.1cm]
            & \quad Info ($n=25$) & Count & 57.0\rlap{\textsuperscript{$\ast\ast$}} & 163.0\rlap{\textsuperscript{$\ast\ast$}} & 36.0\rlap{\textsuperscript{$\ast$}} & 256.0 \\
             & & Expected & 42.4 & 186.5 & 26.9 & 256.0 \\
             & \quad No-Info ($n=31$) & Count & 31.0\rlap{\textsuperscript{$\ast\ast$}} & 224.0\rlap{\textsuperscript{$\ast\ast$}} & 20.0\rlap{\textsuperscript{$\ast$}} & 275.0 \\
             & & Expected & 45.5 & 200.4 & 29.0 & 275.0 \\[.1cm]
			 \cmidrule[0.4pt]{2-7}
			& Reactor \\ [.1cm]
            & \quad Info ($n=27$) & Count & 21.0\rlap{\textsuperscript{$\ast$}} & 185.0\rlap{\textsuperscript{$\ast\ast$}} & 14.0 & 220.0 \\
             & & Expected & 30.3 & 168.4 & 21.2 & 220.0 \\
            & \quad No-Info ($n=21$) & Count & 39.0\rlap{\textsuperscript{$\ast$}} & 148.0\rlap{\textsuperscript{$\ast\ast$}} & 28.0 & 215.0 \\
             & & Expected & 29.6 & 164.5 & 20.7 & 215.0 \\[.1cm]
			\cmidrule[0.4pt]{2-7}
			& Follower \\ [.1cm]
            & \quad Info ($n=8$) & Count & 3.0 & 46.0 & 2.0 & 51.0 \\
             & & Expected & 1.9 & 47.03 & 1.9 & 51.0 \\
            & \quad No-Info ($n=9$) & Count & 1.0 & 49.0 & 2.0 & 52.0 \\
             & & Expected & 2.0 & 47.9 & 2.01 & 52.0 \\[.1cm]
			\cmidrule[0.4pt]{1-7}
			& Total & Count & 152.0 & 815.0 & 102.0 & 1069.0 \\
			 & & Expected & 152.0 & 815.0 & 102.0 & 1069.0 \\			\bottomrule
			\multicolumn{7}{m{12cm}}{\small Counts show the presence of each SA level in players' responses. \newtext{Expected represents the expected counts for the Chi-square test under the null hypothesis (i.e., no association between manipulations and SA levels.)}}\\
            \multicolumn{7}{m{12cm}}{\textsuperscript{$\ast$}\small Chi-square Post-hoc test shows significant association between behavioral profiles or experimental manipulations and SA level at $\alpha=0.05$ with Bonferroni correction.}\\
            \multicolumn{7}{m{12cm}}{\textsuperscript{$\ast\ast$}\small Chi-square Post-hoc test shows significant association between behavioral profiles or experimental manipulations and SA level at $\alpha=0.01$ with Bonferroni correction.}
		\end{tabular}
	}
\end{table}

\subsubsection{Mental Model Development vs. Behavioral Profiles (H3)}
We first investigated the mental model \newtext{development}\oldtext{formation} of players with different behavioral profiles. 
Table~\ref{tab:contingencyTables_study2} shows the frequency of players' comments that reflect each aspect of SA and separated for Hoarders, Reactors, and Followers. 
Chi-square test of independence showed a significant association between behavioral profiles and players' situation model (\newtext{${\chi}^2_{4, 1069}=18.132$, }$p=.001$)\newtext{, and indicated a weak dependence (Cramer's $V=.092$)}. Specifically, Hoarders' comments demonstrate significantly lower comprehension ($p=.005$). According to \newtextfinal{our prior study}~\cite{mohaddesi_trust_2022}, Hoarders over-order more frequently compared to other players. Such behavior is associated with a poor understanding of the system's dynamics, especially when facing a complex decision-making environment~\cite{sterman2015m}:

\begin{quote}
    \textit{``Terrible, the ordering system has made it too confusing to track where the saline is in the system, so I don't know how I can be expected to turn a profit like this''}--(PL2-30, male, 27) \\
    \textit{``Bad. The slow supply of saline has messed things up. I'll have to keep ordering more.''}--(PL2-56, female, 32)\\
\end{quote}

Even if they comprehend their situation as positive, they seem to give more weight to future uncertainties and, as a result, behave proactively. Followers, on the other hand, show significantly higher comprehension ($p<.001$) and lower perception ($p<.001$) in their comments (See Table~\ref{tab:contingencyTables_study2}). The comprehension aspect of the Followers' situation model is mainly reflected in the code ``general-positive'' as exemplified in comments such as ``good'', ``excellent'', ``great'', ``very good''. This could explain why Followers trust the order suggestions and do not deviate: in their mind everything is going well and there is no reason to change. In addition, because they comprehend the state as positive more frequently, they may have less incentive to pay attention to the elements of the system, which can explain their lower perception.  

\begin{figure*}[ht]
    \centering
    \includegraphics[width=.8\columnwidth]{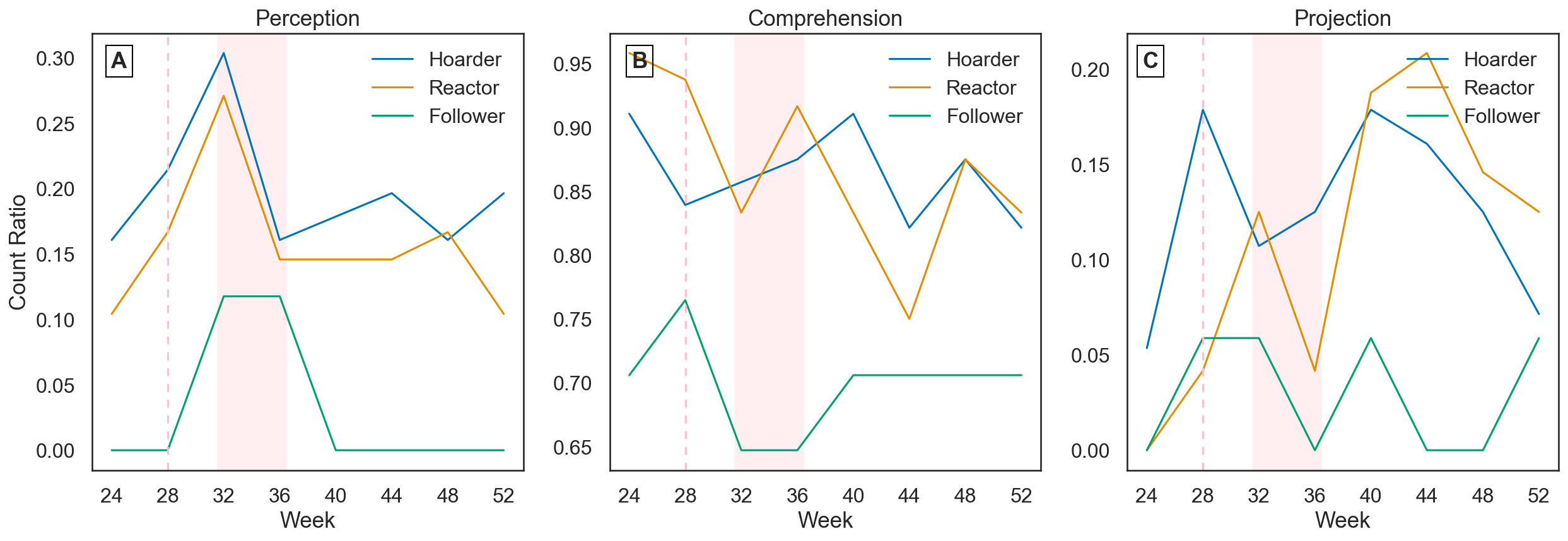}
    \caption{Progression of different aspects of situation model over time depending on behavioral profiles in Study 2. The vertical dashed line in each graph shows the time players were notified about the disruption. The highlighted area illustrates the shortage period due to manufacturing disruption. \newtext{For each behavioral profile group, count ratio was calculated by dividing the number of comments that reflect each SA aspect by total number of players in that group.}
    }\label{fig:study2_sa_over_time_players_only}
    \Description[lineplots showing the average count of SA concepts in Study 2]{three lineplots showing the average count of SA concepts in Study 2 for different behavioral profiles}
\end{figure*}

Next, we examined the progression of mental model \newtext{development}\oldtext{formation} as evidenced by SA levels (Figure~\ref{fig:study2_sa_over_time_players_only}). The disruption seems to have triggered an increase in perception, especially for Hoarders and Reactors, which is later reduced during the shortage period (see Figure~\ref{fig:study2_sa_over_time_players_only}-A). The majority of the comments reflecting perceptions during the disruption period center around ``backlog-increase'' or ``profit-decrease'' codes (see Figure~\ref{fig:study2_sa_topics_over_time_players_only}-A1):

\begin{quote}
    \textit{``The reduced supply from our manufacturers is affecting our supply chain. We could not meet the demands in the past month.''}--(PL2-26, male, 32) \\
    \textit{``We need to get more order to deliver to out customer as we have a backlog we need to supply.''}--(PL2-90, male, not stated) \\
\end{quote}

Hoarders also show a peak in their projection level right after being notified about the disruption on Week 28 (see Figure~\ref{fig:study2_sa_over_time_players_only}-C), which is mainly related to ``anticipating problem/uncertainty'' (see Figure~\ref{fig:study2_sa_topics_over_time_players_only}-C1):

\begin{quote}
    \textit{``So far everything has been smooth, but I anticipate a shortage of inventory due to Covid at our supplier, so I have upped the most recent order to be prepared.''}--(PL2-105, female, 54) \\
    \textit{``Good, ordered more inventory this past week incase [sic] supply chain runs into issues over the next few weeks due to covid.''}--(PL2-39, female, 22)
\end{quote}

As it appears again, Hoarders are more proactive when facing uncertainty. Reactors fall behind in their projections, although showing more projections toward the end (see Figure~\ref{fig:study2_sa_over_time_players_only}-C). Their projections predominantly concern improvement as captured by codes ``general-improve'' and ``profit-improve'' (see Figure~\ref{fig:study2_sa_topics_over_time_players_only}-C2):

\begin{quote}
    \textit{``Badly but hopefully the situation is stabilising''}--(PL2-32, female, 62) \\
    \textit{``The problem will be solved very soon''}--(PL2-68, male, 48)
\end{quote}

Such comments, however, illustrate a lack of clear strategy in Reactors' projections. While they show similar trends in perception levels as Hoarders (Figure~\ref{fig:study2_sa_over_time_players_only}-A), Reactors' mental model development over time does not seem to enable them to be strategic about the future. In sum, these results support H3: we observe that behavioral profiles result in a different mental model \newtext{development}\oldtext{formation}. 





\subsubsection{Mental Model Development vs. Information Sharing per Behavioral Profile (H4)}
\begin{figure*}[ht]
    \centering
    \includegraphics[width=.8\textwidth]{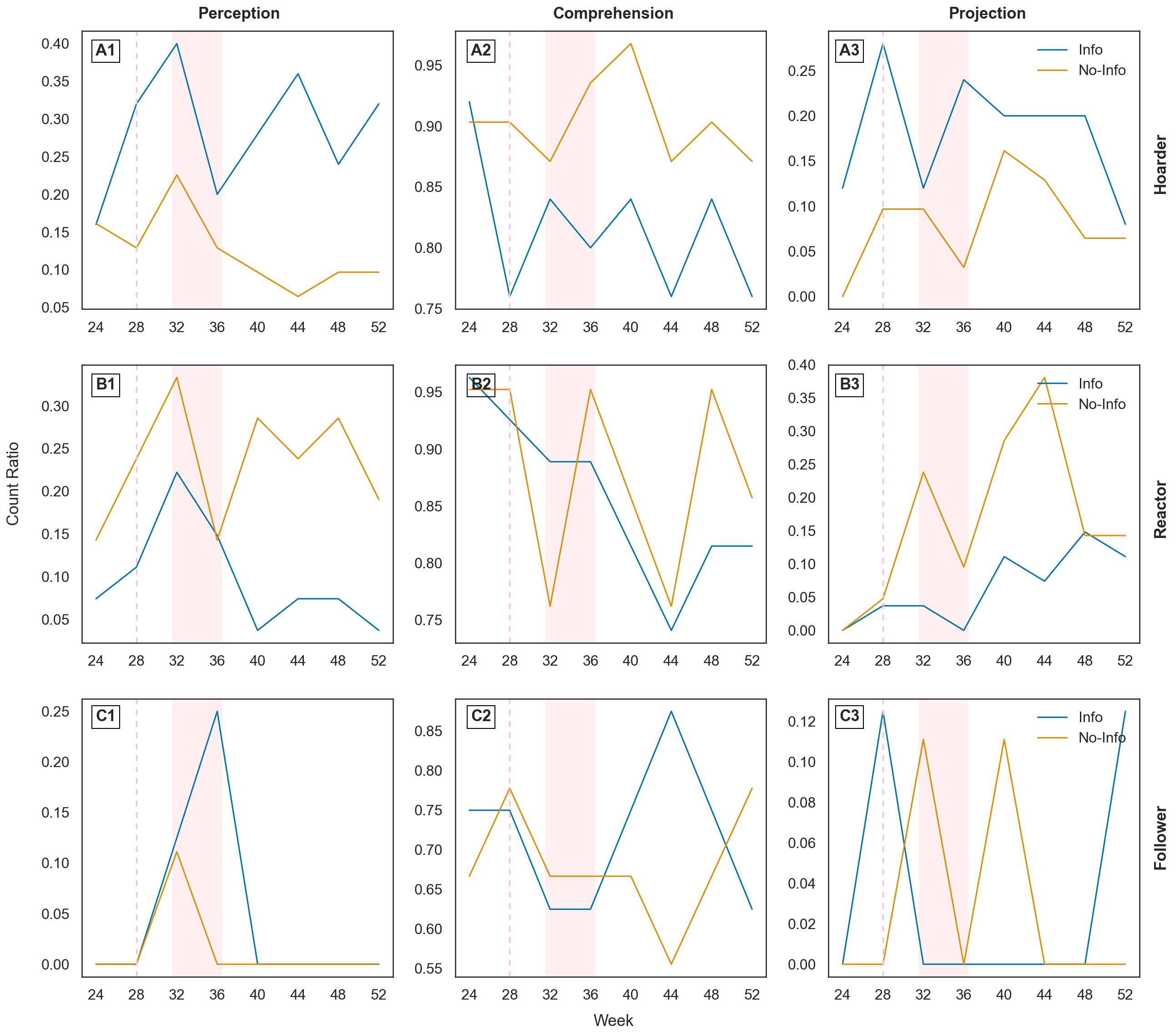}
    \caption{Progression of different aspects of situation model over time depending on behavioral profiles and level of information sharing in Study 2. The vertical dashed line in each graph shows the time players were notified about the disruption. The highlighted area illustrates the shortage period due to manufacturing disruption. \newtext{For each experimental group, count ratio was calculated by dividing the number of comments that reflect each SA aspect by total number of players in that group.}}\label{fig:study2_sa_over_time_players_and_mainpulation}
    \Description[nine lineplots showing the average count of SA concepts in Study 2]{nine lineplots showing the average count of SA concepts in Study 2 for different behavioral profiles and by ploting separate lines for Info and No-Info conditions.}
\end{figure*}

To study how information sharing affects the mental model \newtext{development}\oldtext{formation} of players with different behavioral profiles, we followed the same steps as in previous analyses. First, we counted the presence of each SA aspect in Hoarders' comments, where we found a significant association between information sharing and SA levels (\newtext{${\chi}^2_{2, 531}=21.216$, }$p<.001$)\newtext{, indicated by a small effect (Cramer's $V=.200$)}. According to chi-square post-hoc test (Table~\ref{tab:contingencyTables_study2}), being a Hoarder and having access to information is associated with significantly higher perception ($p<.001$), significantly less comprehension ($p<.001$), and significantly higher projection ($p=.005$). In addition, comments from Hoarders in the Info group show fluctuations of SA aspects over time (see Figure~\ref{fig:study2_sa_over_time_players_and_mainpulation}-A1-3). We can see an increase in the perception aspect of these Hoarders after the disruption notification (i.e., Week 28 - see Figure~\ref{fig:study2_sa_over_time_players_and_mainpulation}-A1). We noticed that comments from Hoarders in the Info group do not directly reflect the perception of the supplier inventory. However, their comments do show signs of becoming more proactive and perhaps using the information on supplier inventory to hoard before shortage affects their supply:

\begin{quote}
    \textit{``Good, ordered more inventory this past week incase [sic] supply chain runs into issues over the next few weeks due to covid.''}--(PL2-39, female, 22)\\
    \textit{``we have as much available as possible heading into uncertain times.''}--(PL2-50, female, 35)
\end{quote}

This pattern is also represented in the projection aspect of Hoarders in the Info group (Figure~\ref{fig:study2_sa_over_time_players_and_mainpulation}-A3). It seems access to information affects their mental model \newtext{development}\oldtext{formation} by adding to their uncertainties:

\begin{quote}
    \textit{``The shutdown of the factory was a suprise [sic]. I'll have to make sure we keep a good supply.''}--(PL2-56, female, 32) \\
    \textit{``Okay, but concerned about future supply issues.''}--(PL2-60, female, 35)
\end{quote}

As mentioned before, poor understanding of the system's dynamics makes hoarding behavior more likely. Hoarders in Study 2 also represent this poor understanding through generic descriptions/adjectives such as ``general-positive'', ``general-negative'', ``general-neutral'' (see Figure~\ref{fig:study2_sa_topics_over_time_players_only}-B2). Providing information to Hoarders does not necessarily increase their understanding of the system. It does, however, seem to reduce their lack of understanding by directing their attention to other aspects (through increased perception) and leveraging information to make larger safety stocks in preparation for future uncertainties (i.e., projection).

We also found a significant association between information sharing and mental model \newtext{development}\oldtext{formation} for Reactors (\newtext{${\chi}^2_{2, 435}=14.122$, }$p<.001$)\newtext{, which was indicated by a small effect (Cramer's $V=.180$)}. In addition, the chi-square post-hoc test revealed that being a Reactor and having access to information is associated with significantly lower perception ($p=.004$) and higher comprehension ($p<.001$)(see Table~\ref{tab:contingencyTables_study2}), which is the exact opposite of Hoarders. While the results for projection is not significant, average counts of SA aspects over time (see Figure~\ref{fig:study2_sa_over_time_players_and_mainpulation}-B1-3), display an increasing trend in the number of comments reflecting a projection for Reactors in both groups (Figure~\ref{fig:study2_sa_over_time_players_and_mainpulation}-B3). Reactors in the Info group show an increase in their perception followed by a decline after the shortage (Figure~\ref{fig:study2_sa_over_time_players_and_mainpulation}-B1). The content of their perceptions mainly reflects ``profit-decrease'' and ``backlog-increase'' (see Figure~\ref{fig:study2_sa_topics_over_time_players_and_mainpulation}-A2):

\begin{quote}
    \textit{``We incured [sic] a lot of losses this month. we are not able to meet up with the demands from the health centers.''}--(PL2-26, male, 32) \\
    \textit{``not great our backlog is large''}--(PL2-4, female, 36)
\end{quote}

While Reactors in the Info group show significantly higher comprehension, the content of their comprehension is mainly summarized in ``general-positive'', ``general-negative'', and ``general-neutral'' codes similar to Hoarders. These codes are represented in comments such as ``good'', ``perfect'', ``fine'', ``bad'', ``Not good'', and ``terrible''. Such comments \textit{can} be again a sign of poor understanding of the system's dynamics. In contrast with Hoarders, it seems information sharing does not direct Reactors' attention to other elements as much (through perception), which affects how they comprehend their state. Reactors in the Info-group also show a declining trend in the counts of their comprehension-related codes over time (Figure~\ref{fig:study2_sa_over_time_players_and_mainpulation}-B2).
Reactors start to hoard when facing the disruption, but it seems that those who have access to their supplier inventory rather wait and see how things unfold instead of being proactive. Both Hoarders and Reactors engage in hoarding behavior as represented in their actions. However, they show different hoarding patterns that seem to originate from different mental model development, which is affected by information sharing.

Finally, the chi-square test of independence did not show a significant association between information sharing and mental model development of Followers (\newtext{${\chi}^2_{2, 103}=1.085$, }$p=.58$\newtext{, Cramer's $V=.103$}). We report the chi-square test result to be consistent with previous analyses. However, we acknowledge that the contingency table for Followers (Table~\ref{tab:contingencyTables_study2}) violates the assumption of the chi-square test by having expected frequencies less than 5. We also performed Fisher's exact test~\cite{field2013discovering}, which similarly revealed no significant association ($p=.65$). It seems information sharing does not affect Followers mental model development. They continue to believe that ordering the same as the system's suggestion is the best strategy. The average counts of SA aspects over time also did not reveal a specific trend (Figure~\ref{fig:study2_sa_over_time_players_and_mainpulation}-C1-3). These results supports our hypothesis (H4) by showing how information sharing is associated with mental model \newtext{development}\oldtext{formation} depending on the behavioral profile. \oldtext{We characterized Hoarders, Reactors, and Followers from the lens of their observed behavior. Our results demonstrate how these behaviors are reflected in payers' mental model development captured via thought bubbles. The results from Mohaddesi et al.
provided evidence for the effect of information sharing on the observed behavior of different player types. Our results help open a window on mental model development over time and how information sharing plays a role in explaining such behaviors.}

\newtext{\subsection{Summary}\label{study2_summary}
Study 2 demonstrated the reliability of our qualitative coding approach for measuring mental model concepts as evidenced by a sufficiently high agreement rate ($\kappa$=$0.84$). We also connected mental model development to player actions by characterizing Hoarders, Reactors, and Followers from the lens of their observed behavior. Our results demonstrated that differences in decision patterns result from differences in mental model development. In particular, Hoarders' proactive behavior is the outcome of more uncertainty about the future as they fail to make sense of the situation. In contrast, the Followers' mental model development directs them to interpret the situation positively and trust the order suggestions. Disruption also affects players' mental model development over time, directing Hoarders' and Reactors' attention to elements of the environment, evidenced by increased perception (see Figure~\ref{fig:study2_sa_over_time_players_only}-A).

Study 2 also complemented Study 1 by providing evidence on how the effect of information sharing on observed behavior is, in fact, the outcome of the mental model development of players over time. More specifically, 
\begin{itemize}
    \item Hoarders exhibit higher perception, lower comprehension, and higher projection when having access to their supplier inventory, leading them to be more proactive and stockpile.
    \item Reactors display lower levels of perception, and higher comprehension, which directs them towards less strategic behavior and merely react to uncertainties.
    \item Followers' mental model development does not seem to be affected by information sharing. 
\end{itemize}
These results are aligned with evidence \newtextfinal{we found in our previous research}\oldtextfinal{provided by Mohaddesi et al.} for the effect of information sharing on the observed behavior of different player types~\cite{mohaddesi_trust_2022}. Using thought bubbles, however, we opened a window on mental model development (as a cognitive construct) over time and demonstrated how information sharing plays a role in explaining such behaviors from a cognitive aspect.}



\section{Discussion and Conclusion}
The results from our studies provided empirical support for the use of thought bubbles. A striking finding was the complete opposite patterns in mental model development of Hoarders and Reactors. Our results not only complemented the behavioral patterns from \newtextfinal{our prior research}\oldtextfinal{Mohaddesi et al.}~\cite{mohaddesi_trust_2022}, but provided a deeper theoretical understanding of the cognitive aspects of human decisions in their interaction with dynamic systems. Understanding cognitive constructs of mental models is instrumental in making sense of human behavior and has been of interest to the HCI field for decades~\cite{hollan2000distributed}. This is an elusive task and requires elicitation of mental models and analyzing elicited concepts. We used thought bubbles for mental model elicitation by evoking players' thinking and capturing verbalization of their thought processes. Of course, many other researchers have also studied elicitation of mental models~\cite{jones2014eliciting,grenier2015conceptual}. Our work is distinguished from these studies in the sense that we attempted to elicit mental models in \newtext{(1)} a diegetic setting---in the context and during the interaction\newtext{, (2) using an open-ended prompt to collect verbalization of thought processes, and (3) over time}.

\newtext{The concept of elicitation is fundamental in studying mental models.}\oldtext{This idea of diegetic elicitation is particularly important considering that mental models are dynamic constructs
. They dynamically form based on new information and as the result of interaction with the environment. Therefore, through diegetic elicitation, we opened up an opportunity to capture thought processes over time, providing a better window into mental model development.} In the HCI context, numerous studies on eliciting mental models exist~\cite{gero2020mental,banks2021some}. However, most of them approach the elicitation process in a non-diegetic form, for example, using interviews before or after the interaction. \newtext{While a comparison with non-diegetic elicitation techniques was out of the scope of our work, we showed how} \oldtext{D}\newtext{d}iegetic elicitation allows for dynamic assessment of mental model development over time\newtext{, and by}\oldtext{while} having minimal interruption to the user.
\newtext{The idea of diegetic elicitation pertains to prior research studying how elicitation procedures affect mental models. Jones et al.~\cite{jones2014eliciting} provided evidence on how the interview process affects mental model representations and found out that out-of-context elicitation pertains to mental models stored in long-term memory. Doyle et al.~\cite{doyle2008measuring} argued that interviewers or additional information after the task can impact subjects' mental models, and thus, subjects should be isolated from any out-of-context influences. Therefore, non-diegetic elicitation may measure mental models that are different from what players relied on during the interaction.}
In addition, considering the context of our study (i.e., dynamic decision-making), it made sense to elicit mental models in situ and diegetically, as in this context people continuously interact with the decision environment which affects their mental model development. We designed thought bubbles as part of a meeting scene (Figure~\ref{fig:thoughtbubbles}) to ensure keeping players in the context.


We chose to elicit mental models through a verbal process, asking players to articulate their thinking. We found inconsistencies in prior research regarding the value of verbal elicitation. \oldtext{While some}\newtext{Some} researchers advocated for the use of textual data and argued that language is key to understanding mental models~\cite{carley1992extracting}\newtext{. }\oldtext{, others}\newtext{Others} have pointed to the complex nature of mental models as a cognitive construct that makes it difficult for individuals to articulate them~\cite{memon2013enhanced}. While we acknowledge that not all aspects of mental models can be verbalized, our results demonstrated that the parts that are verbalized provide much insight, especially about the mental model development of people with different behavioral profiles. We accomplished this by designing thought bubbles with an open-ended prompt to mitigate contextual and framing biases~\cite{memon2013enhanced} and to avoid taking players out of context. Although one might be skeptical of players' engagement with thought bubbles in this format, despite a somewhat negative trend, we found evidence for continuous engagement across the eight prompts (see Appendix~\ref{appendix:word_counts}). A\newtext{n} interesting finding was differences in the average number of words per comment depending on behavioral profile and manipulations. For example, hoarders in the Info group and Reactors in the No-info group wrote more words per comment (see Figure~\ref{fig:study2_word_counts}). 


\newtext{Using thought bubbles, we collected data on players' thought processes over time and in intervals. Mental model elicitation over time is particularly important, considering that mental models are dynamic constructs~\cite{crandall2006working}. They dynamically form based on new information and as the result of interaction with the environment. Therefore, elicitation over time allowed us to consider this dynamic nature and opened up an opportunity to study mental model development as a process. Of course, the way mental model measurement should be conceptualized (i.e., as a process or an outcome) depends on the research question and the underlying task. However, prior studies also seem to advocate for the process view on mental models, pointing to their dynamic nature~\cite{staggers1993mental}, or because of measurement errors arguing that measuring the change in mental models through repeated elicitation is preferred~\cite{doyle2008measuring}. Our results demonstrated the benefit of elicitation over time by showing how experimental manipulations and behavioral profiles influence the mental model development of players over time, as represented by the aspect of their situation model (see Figures~\ref{fig:study1_sa_over_time_players_and_mainpulation}-\ref{fig:study2_sa_over_time_players_and_mainpulation}).}

\newtext{From a methodological perspective, our approach for testing thought bubbles involved three main elements: (1) game environment, (2) dynamic decision-making task, and (3) interaction with a complex dynamic system. Future research can leverage thought bubbles by adapting any of these elements with some considerations. First, while games are particularly useful setting for knowledge elicitation~\cite{van2014system}, we can envision implementing thought bubbles in a non-game environment; for example, as part of a simulation interface where subjects provide textual input (in intervals) on their thought process. However, we argue that the characteristics of simulation games (here gamettes)---including their realism in representing the system, ability to foster communication, and active involvement~\cite{lukosch2018scientific}---can lead players to engage with thought bubbles differently in a game setting compared to a non-game one. 
Future research can study how well a non-game interface can capture thought processes in relation to the task with thought bubbles.
}

\newtext{Second, we studied the use of thought bubbles in a dynamic decision-making task. When studying a different type of task, one must ask if the focus is on measuring mental models as a process or an outcome~\cite{staggers1993mental}. If the focus is on the mental model representation and their content or structure in relation to the task---while we advocate leveraging thought bubbles for diegetic elicitation---querying subjects in intervals might be less relevant. 
Although considering the studies that cast doubt on the externalization of mental models as a static construct~\cite{doyle2008measuring}, and prefer the process view on mental models~\cite{staggers1993mental}, future research must scrutinize to what extent eliciting mental model development as a process can be justified for other types of tasks. Another key consideration is whether the goal is to understand mental models or improve the mental models of the task. While improving mental models was not the purpose of our study, we showed how experimental manipulations such as information sharing affect mental model development. Therefore, future research should look into how manipulations can improve mental models and enhance performance, perhaps by incorporating smart nudging interventions~\cite{mele2021smart}. Another way is to leverage AI for different framing of the prompt, especially as prior research argued how wording and framing of questions could affect mental models~\cite{doyle2008measuring}. 

The type of task also affects the choice of framework for analyzing elicitation results. We used the Situation Awareness model introduced by Endsley~\cite{endsley1995toward} as it maps well to mental models theory and dynamic decision-making context}\oldtext{In making sense of elicited concepts through qualitative analysis, we found that the Situation Awareness model introduced by Endsley
is particularly suitable. The SA model helped us in our analysis as it maps well to mental models theory and provided us with a consistent framework for gaining a theoretical understanding of players' mental model development over time. In addition, SA was introduced in the dynamic decision-making context
}, and hence, suited our study in which participants interacted with a dynamic \newtext{decision task}\oldtext{system}. However, when \newtext{the task is}\oldtext{interactions are} not dynamic \newtext{decision-making} (e.g., \newtext{gestural} interaction with a display~\cite{soni2020adults}), SA might be less relevant. While we advocate for grounding the qualitative analysis process in theories, the choice of underlying theoretical framework in non-dynamic context requires more scrutiny. We demonstrated the reliability of our approach by using our generated codebook from Study 1 to a new dataset (Study 2) and providing insight on mental model development of players with distinct behavioral profiles.

Finally, \newtext{we studied players' mental model development in their interaction with an interactive environment simulating a complex dynamic system (i.e., supply chain simulation that evolved both spontaneously and as the result of players' actions).}\oldtext{we used games to elicit mental models of decision-makers in a dynamic environment. Prior research argues that games are particularly useful for eliciting knowledge
.} However, HCI encompasses many forms of interaction~\cite{helander2014handbook}. Future research can study other \newtext{types}\oldtext{forms} of interaction in which thought bubbles can be leveraged\newtext{, perhaps with some adaptions. For example, researchers can investigate the use of thought bubbles for diegetic elicitation through non-verbal queries such as card sorting
~\cite{martin2022supporting}, or diagrammatic representations
~\cite{saint2021diagrammatic}. In addition}\oldtext{. For example}, recent advances in AI have inspired many researchers to study human-AI interaction~\cite{amershi2019guidelines, sundar2020rise} and investigate mental models of AI~\cite{gero2020mental, villareale2022want}. Thought bubbles can be utilized to advance our understanding of users' mental model development of AI over time. Thought bubbles can also be viewed as a form of reflection, and reflection can be used to improve learning~\cite{helyer2015learning}. Therefore, future research can look into opportunities in which thought bubbles can be utilized for reflection in game-based learning technologies~\cite{harteveld2020gaming4all}. For all this to work, we need to think about how to scale up the use of thought bubbles with respect to qualitative analysis, which is a time-consuming process. Future studies can look into leveraging natural language processing (NLP) for qualitative analysis~\cite{leeson2019natural, guetterman2018augmenting, crowston2012using}. Last but not least, thought bubbles can potentially elicit more than mental models of the tasks at hand by uncovering cognitive aspects such as attitudes and motivation, which is another interesting future avenue, as reflected in the following quote:

\begin{quote}
``This is bad. I guess I was scared about having enough supply. I over ordered and now we are paying for it in surplus inventory. I should have gone to veterinarian school like my family wanted! But no! I was all like ''I'm going to be a great supply chain manager one day''''
\end{quote}

\begin{acks}
This research was supported with funding from the National Science Foundation (NSF: 1638302 \& 2028449). We further thank the StudyCrafter and Drug Shortage teams at Northeastern University.
\end{acks}

\balance
\bibliographystyle{ACM-Reference-Format}
\bibliography{references}

\appendix


\newpage
\onecolumn
\section{Distribution of Qualitative Codes Over Time for Behavioral Profiles}\label{appendix:study2_sa_topics_over_time_players}

\begin{figure*}[ht]
    \centering
    \includegraphics[width=\textwidth]{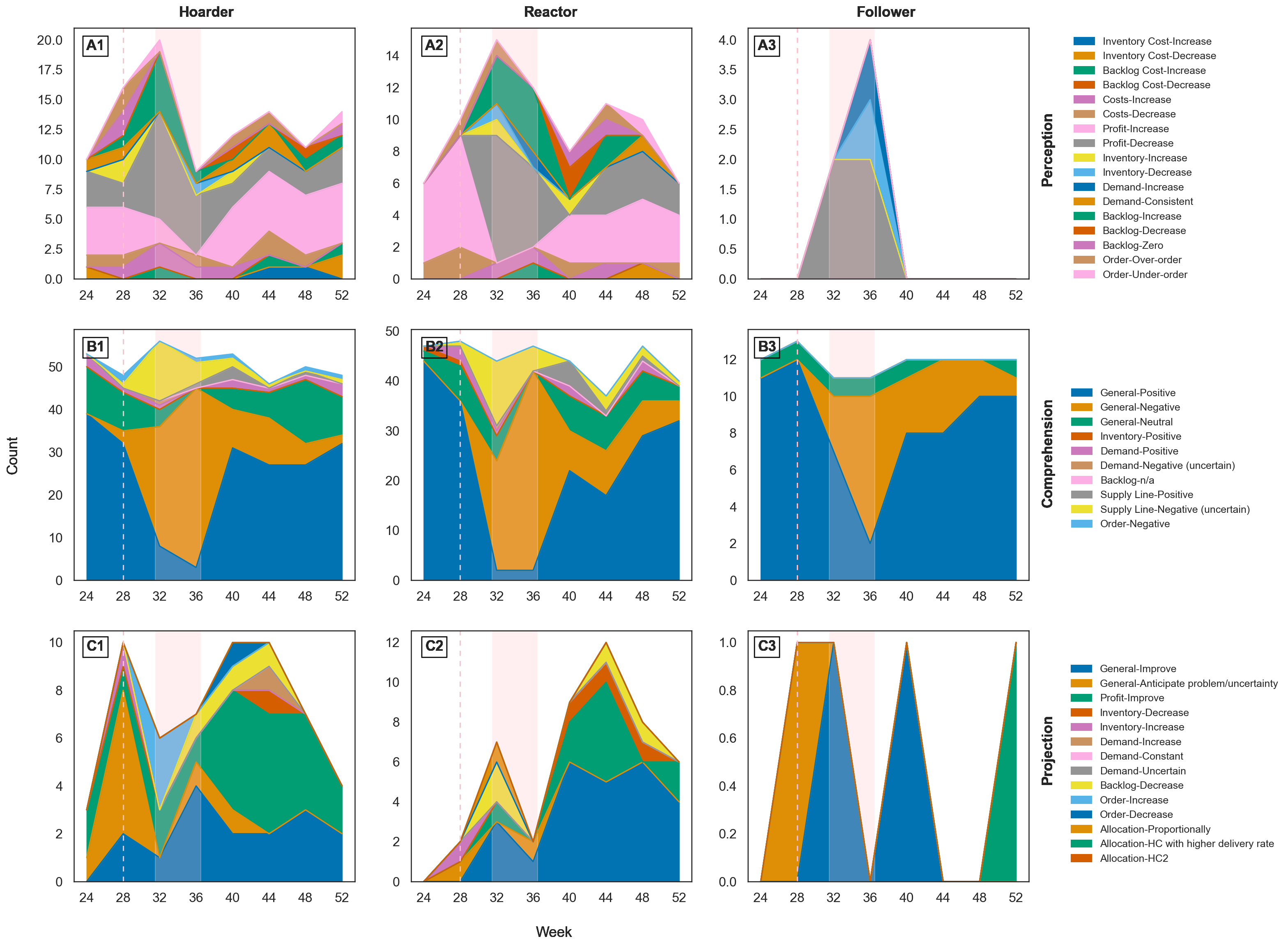}
    \caption{Distribution of topics and descriptions from qualitative coding over time depending on behavioral profiles in Study 2. The vertical dashed line in each graph shows the time players were notified about the disruption. The highlighted area illustrates the shortage period due to manufacturing disruption.
    }\label{fig:study2_sa_topics_over_time_players_only}
    \Description[Distribution of topics and descriptions from qualitative coding over time]{nine area plots showing the count of qualitative topics and descriptions for players in Study 2 for different behavioral profiles across perception, comprehension and projection.}
\end{figure*}

\newpage
\section{Distribution of Qualitative Codes Over Time for Behavioral Profiles and Depending on Information Sharing}\label{appendix:study2_sa_topics_over_time_players_and_info}

\begin{figure*}[ht]
    \centering
    \includegraphics[width=\textwidth]{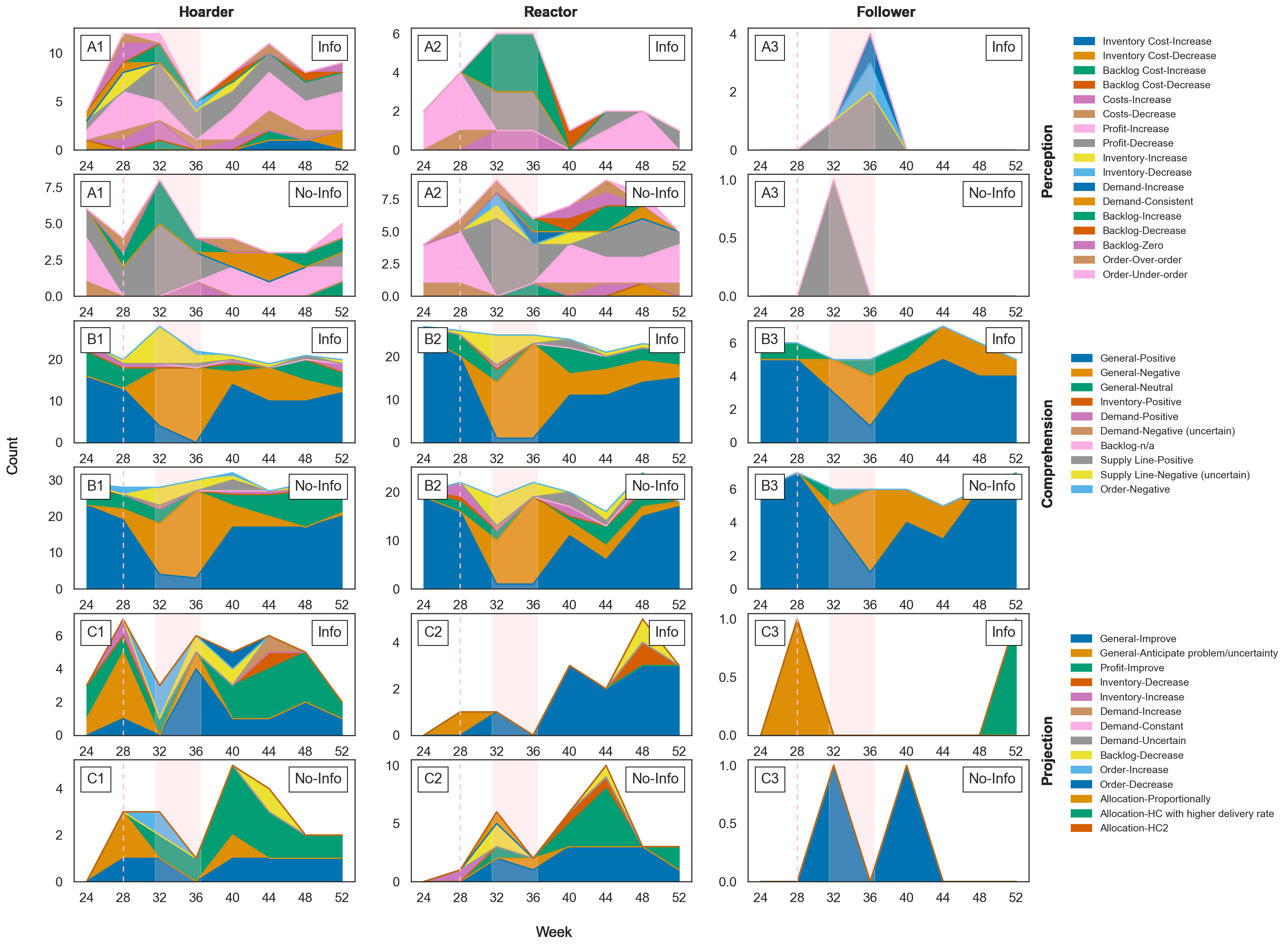}
    \caption{Distribution of topics and descriptions from qualitative coding over time depending on behavioral profiles and level of information sharing in Study 2. The vertical dashed line in each graph shows the time players were notified about the disruption. The highlighted area illustrates the shortage period due to manufacturing disruption.
    }\label{fig:study2_sa_topics_over_time_players_and_mainpulation}
    \Description[Distribution of topics and descriptions from qualitative coding over time for information sharing]{18 area plots showing the count of qualitative topics and descriptions for players in Study 2 for different behavioral profiles and levels of information sharing across perception, comprehension and projection.}
\end{figure*}

\newpage
\section{Average Count of Word per Comments}\label{appendix:word_counts}

\begin{figure*}[ht]
    \centering
    \includegraphics[width=.8\textwidth]{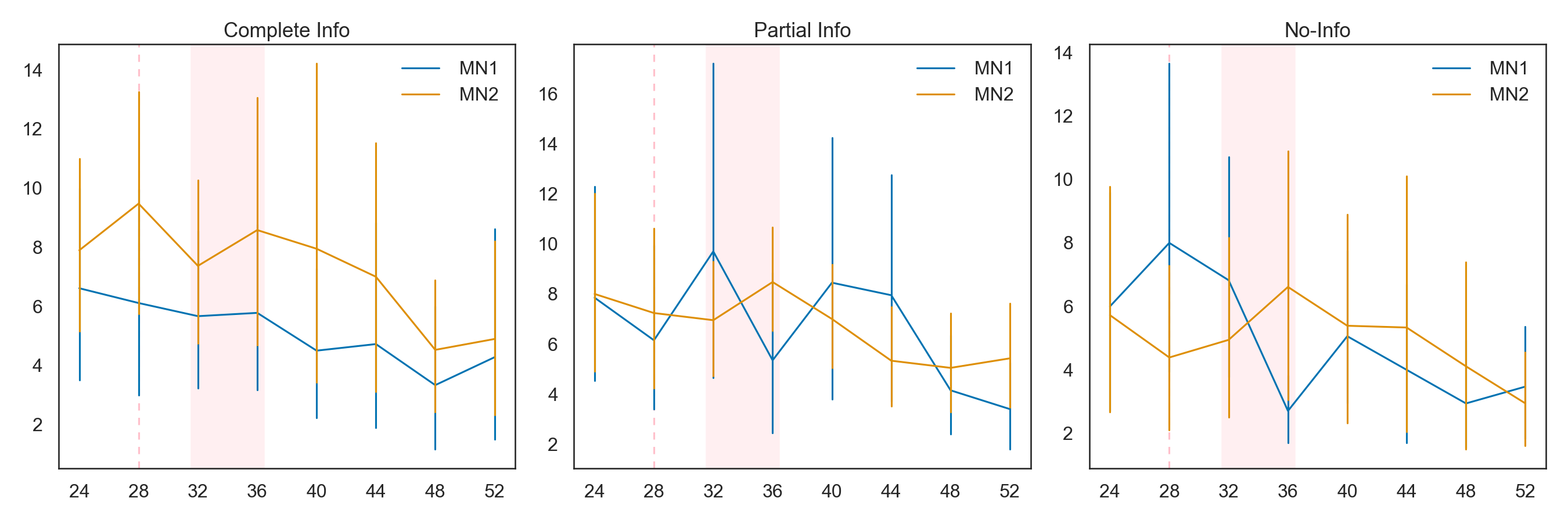}
    \caption{Average count of words per comment for players in Study 1. The vertical dashed line in each graph shows the time players were notified about the disruption. The highlighted area illustrates the shortage period due to manufacturing disruption.
    }\label{fig:study1_word_counts}
    \Description[lineplots of word counts for Study 1]{three separate lineplots showing the average count of words per comment for players in Study 1}
\end{figure*}

\begin{figure*}[ht]
    \centering
    \includegraphics[width=.8\textwidth]{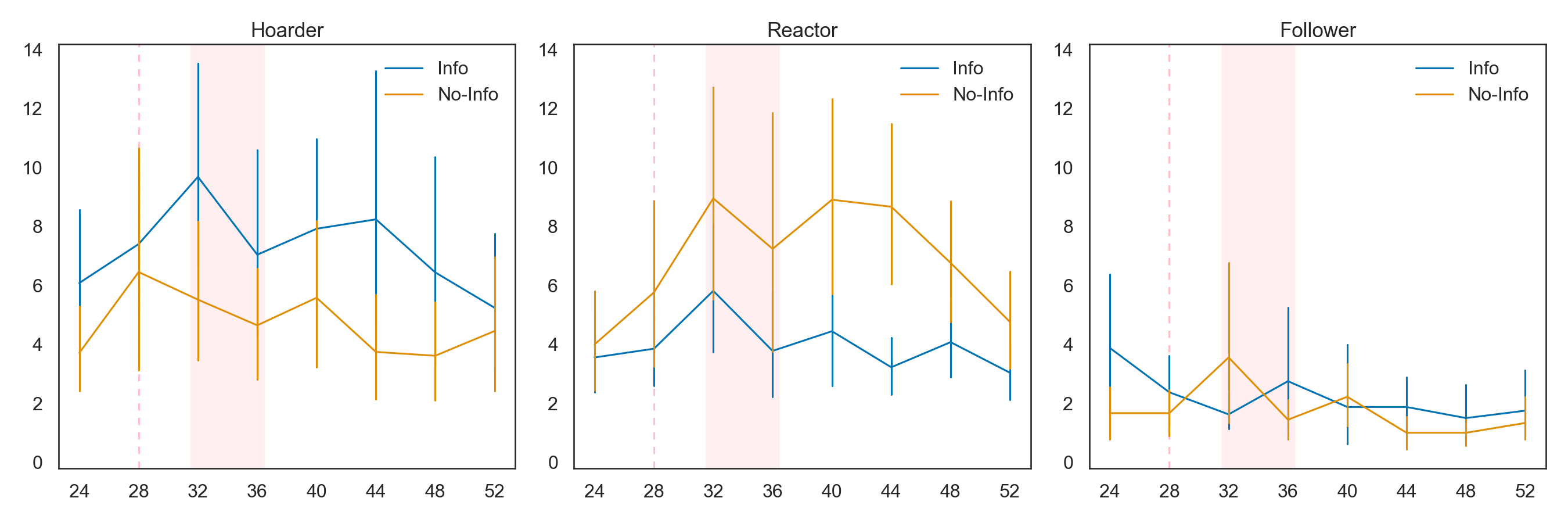}
    \caption{Average count of words per comment for players in Study 2. The vertical dashed line in each graph shows the time players were notified about the disruption. The highlighted area illustrates the shortage period due to manufacturing disruption.
    }\label{fig:study2_word_counts}
    \Description[lineplots of word counts for Study 2]{three separate lineplots showing the average count of words per comment for players in Study 2}
\end{figure*}

\end{document}